\documentclass[twocolumn,usenatbib]{mnras}

\usepackage{subcaption}
\usepackage{graphicx}
\usepackage{amsmath}
\usepackage{siunitx}
\usepackage{physics}
\usepackage{xfrac}
\usepackage{xspace}
\usepackage{comment}
\usepackage{hyperref}
\usepackage{amssymb}

\def\pmb#1{\setbox0=\hbox{#1}%
    \kern-.025em\copy0\kern-\wd0
    \kern.05em\copy0\kern-\wd0
    \kern-.025em\raise.0433em\box0}

\def\ltsima{$\; \buildrel < \over \sim \;$}
\def\gtsima{$\; \buildrel > \over \sim \;$}
\def\simlt{\lower.5ex\hbox{\ltsima}}
\def\simgt{\lower.5ex\hbox{\gtsima}}
\def\p2Y{\;_2Y}
\def\m2Y{\;_{-2}Y}

\def\mk2{\mu {\rm K}^2}
\def\Planck{\it Planck \rm}

\def\LCDM{$\Lambda{\rm CDM}$ }

\def\LCDMns{$\Lambda{\rm CDM}$}

\def\pmb#1{\setbox0=\hbox{#1}%
     \kern-.025em\copy0\kern-\wd0
     \kern.05em\copy0\kern-\wd0
     \kern-.025em\raise.0433em\box0}

\usepackage{xcolor}

\definecolor{purple}{RGB}{156,81,182}

\begin{document}

\title[Disentangling dark matter with weak lensing]{Prospects for disentangling dark matter with weak lensing}

\author[Preston, Rogers, Amon \& Efstathiou]{Calvin Preston$^{1,2}$\thanks{E-mail: cp662@cam.ac.uk}, Keir K. Rogers$^{3,4}$\thanks{E-mail: k.rogers24@imperial.ac.uk}, Alexandra Amon$^{5}$\thanks{E-mail: alexandra.amon@princeton.edu}, George Efstathiou$^{1,2}$\thanks{E-mail: gpe@ast.cam.ac.uk} \\
${1}$ Kavli Institute for Cosmology Cambridge, University of Cambridge,
Madingley Road, Cambridge, CB3 0HA, United Kingdom \\
${2}$ Institute of Astronomy, University of Cambridge, Madingley Road, Cambridge, CB3 0HA. United Kingdom \\
${3}$ Department of Physics, Imperial College London, Blackett Laboratory, Prince Consort Road, London, SW7 2AZ, United Kingdom \\
${4}$ Dunlap Institute for Astronomy and Astrophysics, University of Toronto, 50 St. George Street, Toronto, ON M5S 3H4, Canada \\
${5}$ Department of Astrophysical Sciences, Princeton University, Peyton Hall, Princeton, NJ 08544, United States of America}

\maketitle
\begin{abstract}
We investigate the degeneracy between the effects of ultra-light axion dark matter and baryonic feedback in suppressing the matter power spectrum. We forecast that galaxy shear data from the Rubin Observatory’s Legacy Survey of Space and Time (LSST) could limit an axion of mass $m = 10^{-25}\,\mathrm{eV}$ to be $\lesssim 5\%$ of the dark matter, stronger than any current bound, if the interplay between axions and feedback is accurately modelled. Using a halo model emulator to construct power spectra for mixed cold and axion dark matter cosmologies, including baryonic effects, we find that galaxy shear is sensitive to axions from $10^{-27}\,\mathrm{eV}$ to $10^{-21}\,\mathrm{eV}$, with the capacity to set competitive bounds across much of this range. For axions with $m \sim 10^{-25}\,\mathrm{eV}$, the scales at which axions and feedback impact structure formation are similar, introducing a parameter degeneracy. We find that, with an external feedback constraint, we can break the degeneracy and constrain the axion transfer function, such that LSST could detect a $10^{-25}\,\mathrm{eV}$ axion comprising 10\% of the dark matter at $\sim 3 \sigma$ significance. Direct reconstruction of the non-linear matter power spectrum provides an alternative way of analysing weak lensing surveys, with the advantage of identifying the scale-dependent features in the data that the dark matter model imposes. We advocate for dedicated cosmological hydrodynamical simulations with an axion dark matter component so that upcoming galaxy and cosmic microwave background lensing surveys can disentangle the dark matter-baryon transfer function.

\end{abstract}

\begin{keywords}
cosmology: cosmological parameters, power spectrum, weak lensing, observations, dark matter
\end{keywords}

\section{Introduction}\label{sec:intro}
The cold dark matter (CDM) model with a positive cosmological constant $\Lambda$ (\LCDMns) precisely describes a wide range of cosmological observables. The anisotropies of the cosmic microwave background \citep[CMB,][]{Planck_legacy}, baryonic acoustic oscillations \citep[BAO,][]{Alam:2021a, DESI_BAO_2024, DESI_BAO_DR2}\footnote{The Dark Energy Spectroscopic Instrument (DESI) data release 2 (DR2) has a $\sim 2.0\sigma$ difference in \LCDM parameter inference between CMB data and its BAO measurements.} and many other observables agree remarkably with the predictions of the \LCDM model \citep[see e.g.][]{Planck_legacy}. Nevertheless, the nature of the dark matter particle (or particles) remains unknown, alongside the physical origin of dark energy.

Since the first evidence for the existence of dark halos hosting galaxies was presented by \citet{Rubin_DM}, many candidates for the dark matter have been proposed. Weakly interacting massive particles (WIMPs, most likely stable supersymmetric relic particles) are a leading dark matter candidate \citep[see e.g. the review by][]{Roszkowski_2018}. However, lack of evidence for supersymmetry at the Large Hadron Collider \citep[see e.g.][]{Kadan:2024} and the strong limits on the WIMP cross section from direct detection experiments \citep{WIMP_SNOWMASS} have stimulated interest in other candidates. These candidates include warm, interacting and self-interacting dark matter; complex dark sectors like atomic dark matter; and primordial black holes \citep{Khlopov_1985} and massive compact halo objects \citep[for reviews, see e.g.][]{DM_DE_review,LSST_DM_summary,Green_PBH_review}.

There is renewed interest in axion, or axion-like, particles as candidates for dark matter. The axion was first proposed to solve the strong CP problem in quantum chromodynamics and was soon identified as an excellent dark matter candidate \citep{Peccei_strongCPaxion, Weinberg_axion, Wilczek_1978, Abbott_invisibleaxion, Dine_Fischler, Preskill_axion}. Axion-like particles also feature in extensions of the Standard Model such as string theories \citep{Witten_string_axion} and are viable dark matter candidates \citep{Hu:2000,Antypas_2022}. Ultra-light axions (ULAs) with masses $10^{-33}\,\mathrm{eV} \leq m_{\rm{ax}} \leq 10^{-18}$eV (often referred to as fuzzy dark matter, FDM) are particularly interesting, as this mass scale is preferred in some ``axiverse'' string theories \citep{Arvanitaki_2010,Hui:2017, Visinelli_2019,Sheridan_2024}.\footnote{In this work, we refer to ULAs as a particle physics motivation for FDM; our results apply to any ultra-light scalar DM with a potential quadratic in the field.} The existence of certain axion-like particles has been considered as a solution to the Hubble tension: the discrepancy between local distance ladder measurements \citep{reiss2022} of the Hubble parameter today $H_{0}$ and the value inferred from the CMB assuming the \LCDM model \citep{Params:2018}. Such early dark energy models \citep[for a review, see e.g.][]{Poulin_EDE} are disfavoured as solutions to the Hubble tension \citep{Efstathiou_EDE}. The Hubble tension will not be discussed further in this work \citep[for more information, see e.g.][]{Efstathiou_RS}.

The low masses of the axions that we consider \((m_\mathrm{ax} \leq 10^{-20}\,\mathrm{eV})\) result in large de Broglie wavelengths (kpc to Mpc), with quantum pressure thus acting on astrophysical scales counteracting gravitational collapse. This wave-like nature of FDM will affect cosmological structure formation \citep[for a review of axions in cosmology, see e.g.][]{Marsh_axion}. In cosmological models with ULAs, the linear growth of structure is suppressed for wavenumbers larger than a Jeans wavenumber \(k_\mathrm{J}\) compared to models in which the dark matter is composed only of standard CDM \citep{Amendola:2005ad,Marsh_Ferreira_2010,Mocz_2015}. The Jeans wavenumber
\begin{equation}\label{equ:jeans_scale}
    k_{\rm{J}} = 2.10\,(1 + z)^{-1/4} \left(\frac{\Omega_{\rm{ax}}h^{2}}{0.12}\right)^{1/4} \left(\frac{m_{\rm{ax}}}{10^{-25}\rm{eV}}\right)^{1/2} \rm{Mpc}^{-1}, 
\end{equation}
where \(\Omega_\mathrm{ax}\) is the axion energy density today, \(z\) is redshift and $h = H_0 / (100 \ {\rm km}\,{\rm s}^{-1}\,{\rm Mpc}^{-1})$. In particular, for \(m_\mathrm{ax} \sim 10^{-25}\,\mathrm{eV}\), the Jeans suppression length is \(\sim 8\,h^{-1}\,\mathrm{Mpc}\).

This scale is of interest for weak lensing measurements and, as proposed in \citet{Rogers_S8tensionaxions}, for the so-called $S_{8}$ tension\footnote{$S_8 \equiv \sigma_8 (\Omega_{\rm m}/0.3)^{0.5}$, where $\Omega_{\rm m}$ is the total matter density today and $\sigma_8$ is the root mean square amplitude of the linear matter fluctuation spectrum today in spheres of radius $8\,h^{-1}\,{\rm Mpc}$.}, where some measurements of the linear clustering amplitude $S_{8}$ \citep[e.g.][]{Heymans:2021, Amon:2021,Secco:2022,dalal2023hyper,li2023hyper, KiDSDES} lie $2\sigma-3\sigma$ lower than the \LCDM value inferred from the \Planck CMB temperature, polarisation and lensing analysis \citep{Params:2018}.\footnote{However, some more recent weak lensing analyses have found $S_{8}$ more consistent with \Planck \citep{KIDS_LEGACY}.} 

\citet{AAGPE2022,CPAAGPE} suggest that either baryonic feedback power suppression or non-standard dark matter physics is a source of bias to low values of $S_{8}$. A scale-dependent suppression in the linear matter power spectrum arising from ULAs would mean that \(S_8\) really is low \citep{Rogers_S8tensionaxions}. The ability to model all scales in weak lensing measurements is therefore important to help identify the nature of dark matter.

Current observational limits on ULAs constrain a combination of axion mass \(m_\mathrm{ax}\) and their contribution to the total energy density \(\Omega_\mathrm{ax}\) (or, equivalently, the fraction of dark matter comprised of axions \(f_\mathrm{ax}\)). Multiple astrophysical phenomena have been used or proposed in constraining ULAs, e.g., the CMB \citep{Hlozek_CMB_axions2015,Hlozek_CMB_axions2018,Rogers_S8tensionaxions}, galaxy clustering \citep{Lague_axions_galaxyclustering,Rogers_S8tensionaxions}, galaxy weak lensing \citep{Dentler_Axions_weaklensing,Kunkel_FDM_lensing}, the Lyman-$\alpha$ forest \citep{Irsic_Lyman_DM, Kobayashi_Lyman_DM,Rogers:2020cup,Rogers_Peiris_Lyman,Rogers_Poulin}, Milky Way substructure \citep{Banik:2019smi,2021PhRvL.126i1101N}, strong gravitational lensing \citep{Laroche_etal}, the UV luminosity function of high-redshift galaxies \citep{Bozek,Winch_Rogers_etal}, pulsar timing arrays \citep{Khmelnitsky:2013lxt,Servant_PTA_axions,EuropeanPulsarTimingArray:2023egv}, FDM effects in star clusters and dwarf galaxies \citep{Marsh_Niemeyer,Dalal_2022}, the kinetic Sunyaev-Zel'dovich effect \citep{Farren_axions} and $21$ cm emission \citep{Bauer2021,Hotinli_2022}. Searches for axion Jeans suppression in the Lyman-$\alpha$ forest rule out a single axion being all the dark matter such that $m_{\rm{ax}}> 2\times 10^{-20}\,\mathrm{eV}$ at 95\% c.l. \citep{Rogers_Peiris_Lyman}. However, axions contributing a fraction (as motivated, e.g., by the axiverse), e.g., $f_\mathrm{ax}<22\%$ of the DM energy density for \(10^{-26}\,\mathrm{eV} \leq m_\mathrm{ax} \leq 10^{-23}\,\mathrm{eV}\) \citep{Winch_Rogers_etal}, are still unconstrained.\footnote{Axions with $m_{\rm{ax}}<10^{-27}\,\mathrm{eV}$ also manifest dark energy properties for some time after matter-radiation equality and will not be considered further in this work \citep[see e.g.][]{Rogers_S8tensionaxions}.}

To test for axion dark matter across the ultra-light mass range, we can use the matter power spectrum from linear \citep{Hlozek_CMB_axions2015,Rogers_S8tensionaxions,Rogers_Poulin} to non-linear (wavenumbers $k \gtrsim 0.2\,h\,\rm{Mpc}^{-1}$) scales across multiple cosmic epochs \citep{Kobayashi_Lyman_DM,Rogers_Peiris_Lyman,Dentler_Axions_weaklensing,Winch_Rogers_etal}. In this work, we consider cosmic shear as a low-redshift (\(z \lesssim 2\)) probe of the matter power spectrum, following the first weak lensing test of ULAs \citep{Dentler_Axions_weaklensing}. \citet{Preston_2024} propose that forthcoming weak lensing surveys \citep[e.g., Vera C. Rubin Observatory, \textit{Euclid}, Nancy Grace Roman Space Telescope;][]{LSSTScience,EuclidForecast,Roman_forecast} should allow accurate measurements of the matter power spectrum from linear to highly non-linear scales. If this can be achieved, then it could be possible to detect the distinctive axion non-linearities with weak lensing surveys. Further, there is a complementarity between searching for axion Jeans suppression in the power spectrum and explicitly fuzzy phenomena like halo cores, interference fringes and oscillating dark matter granules that are harder to model \citep{Marsh_Niemeyer,May_Springel_FDM,Dalal_2022, Lague, Zimmerman_FDM}.

However, in addition to axions, baryonic feedback processes (where gas is ejected from the central regions of galaxy groups and clusters to Mpc scales by active galactic nuclei and supernovae) also suppress the matter power spectrum \citep{BAHAMAS,Springel:2018,Chisari_2019}. Whilst the details of baryonic feedback remain uncertain \citep{vandaalen:2020}, the processes are expected to change the matter power spectrum for $k \gtrsim 0.2\,h\,\rm{Mpc}^{-1}$ in comparison to a feedback-free \LCDM cosmology.

\begin{figure*}
    \begin{minipage}{0.487\textwidth}
    \includegraphics[width=1.\columnwidth]{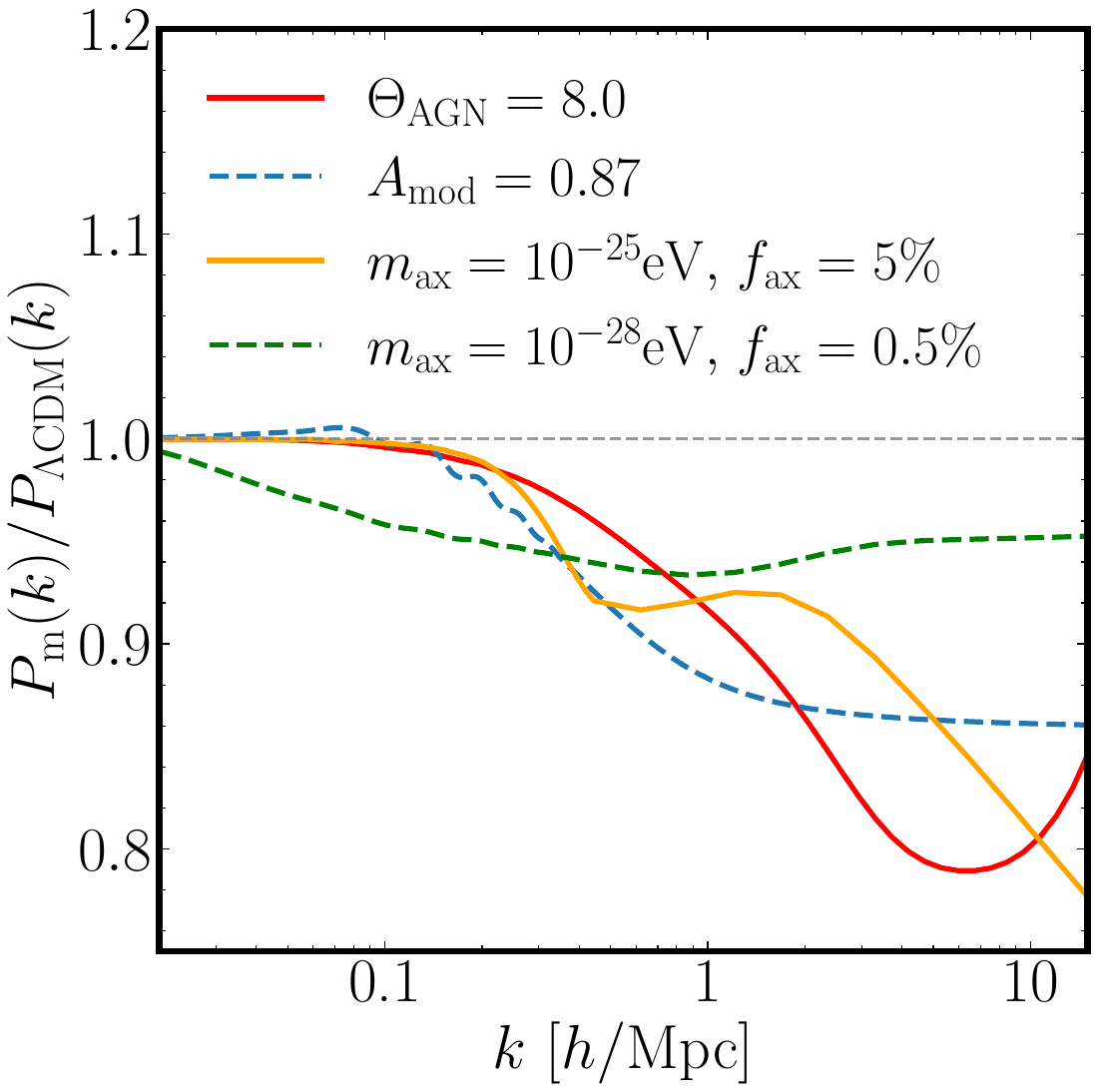} 
    \end{minipage}
    \begin{minipage}{0.499\textwidth}
        \includegraphics[width=1.\columnwidth]{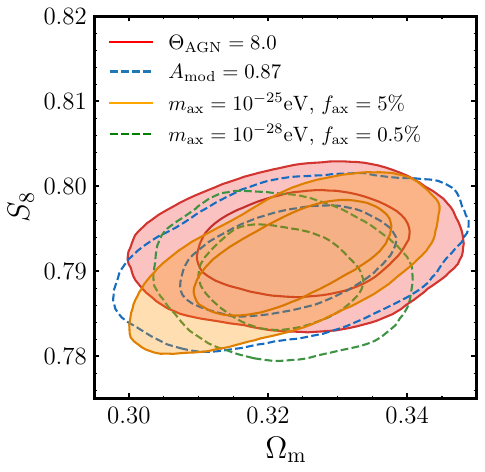} 
    \end{minipage}
\caption{For four cosmological scenarios we show \textit{Left}: the ratio of the power spectrum, $P_\mathrm{m}(k)$, compared to a $\Lambda$CDM dark matter-only case, $P_\mathrm{\Lambda\mathrm{CDM}}(k)$, as a function of wavenumber \(k\) and \textit{Right}: the corresponding forecasted result of a \LCDM inference (ignoring the presence of axions and baryonic feedback) of LSST year 1 cosmic shear data simulated with the four power spectrum scenarios.
The four cases we consider are: strong baryonic feedback from the BAHAMAS simulations $\Theta_{\rm{AGN}} = 8.0$ \citep[red,][]{BAHAMAS}, a phenomenological $A_{\rm{mod}}$ suppression \citep[blue, parameter value chosen from the analysis of the Dark Energy Survey cosmic shear data,][]{AAGPE2022, CPAAGPE} and two axion models \citep{vogt} with masses, \(m_\mathrm{ax}\), each comprising a fraction, \(f_\mathrm{ax}\), of the total dark matter \citep[yellow and green, parameter values chosen to reproduce the low \(S_8\) as proposed in][]{Rogers_S8tensionaxions}. The marginalised posteriors in the $S_{8}-\Omega_{\rm{m}}$ plane are statistically indistinguishable despite the different power spectrum transfer functions; the inner and outer contours respectively indicate the 68\% and 95\% credible regions. This highlights how scenarios with distinct power spectra are degenerate in the $S_8-\Omega_{\rm{m}}$ plane, but would be distinguishable if we could reconstruct their matter power spectra and isolate the differences in shapes.}

\label{fig:toy_example}
\end{figure*}

Since both axions and baryonic feedback suppress the small-scale matter power spectrum, the question arises of whether these effects are exactly degenerate, or whether there is a region of parameter space in which the effects of axions and baryonic feedback can be disentangled. In Fig.~\ref{fig:toy_example}, we show four different models of small-scale power suppression. We show axion models with $m_{\rm ax} = 10^{-25}\,\mathrm{eV}$ (solid yellow curve) and $10^{-28}\,{\rm eV}$ (dashed green curve) respectively comprising a fraction $f_{\rm ax}=5\%$ and $0.5\%$ of the total dark matter density (the rest is CDM); these axion models were proposed by \cite{Rogers_S8tensionaxions} to address the \(S_8\) discrepancy. The dashed blue curve shows the phenomenological model of power suppression introduced by \cite{AAGPE2022} with $A_{\rm mod} = 0.87$; the solid red curve shows the suppression caused by strong baryonic feedback in a standard \LCDM cosmology in the BAHAMAS hydrodynamical simulations \citep[][effective AGN temperature parameter \(\Theta_\mathrm{AGN} = 8.0\)]{BAHAMAS}. Axions are changing both the underlying linear matter power spectrum (initial conditions, scale-dependent growth rate) and the mapping from the linear to the non-linear matter power (axion halo model, fuzzy effects); baryonic feedback is changing only the mapping from the linear to non-linear matter power. A consequence of this difference is that the matter power spectrum at earlier times and in different environments will be affected differently \citep[e.g.,][]{Winch_Rogers_etal,Rogers_Poulin}.

Following \citet{Preston_2024}, we construct shear correlation functions $\xi_{+/-}(\theta)$ (\(\theta\) is angular separation) forecast for the \textit{Rubin} Legacy Survey of Space and Time (LSST) year 1 (Y1) data release (see Appendix \ref{sec:app_A}). The right-hand panel of Fig.~\ref{fig:toy_example} shows forecast LSST Y1 posteriors in the $S_{8}-\Omega_{m}$ plane for the true cosmologies with axions and feedback introduced above, but with axions and feedback ignored in the model that is inferred (i.e. we infer using a standard \LCDM model without feedback). In this analysis, we vary the \LCDM cosmological and shear nuisance parameters listed in Table \ref{tab:priors}, therefore marginalising uncertainties in galaxy intrinsic alignments, the source redshift distribution, shear calibration and cosmological parameters (but not axion and feedback parameters). We use all scales of the shear data vector down to a minimum angular separation of $\sim 2.5$~arcmin (see Fig.~\ref{fig:correlation_functions} for more details).

The models shown in Fig.~\ref{fig:toy_example} are chosen so that the $S_8-\Omega_m$ posteriors are statistically indistinguishable. Each posterior is biased from the true $S_{8}$ value (which is already lower than the equivalent \LCDM value in the axion case), thus displaying an apparent ``$S_8$ tension''. The posteriors peak at $S_8 \approx 0.79$; for the BAHAMAS and $A_{\rm{mod}}$ models, the true $S_8$ is set to the \Planck value of $0.83$, for the axion models, we set $S_{8}\approx0.81$. Whilst these models are indistinguishable in $S_{8}$, they could in principle be distinguished by reconstructing the matter power spectrum.

In this work, we investigate the interplay between axions and baryonic feedback in weak lensing and answer three main questions:
\begin{enumerate}
\item What is the combined effect of axions and baryonic feedback in weak lensing?
\item What is the sensitivity of LSST Y1 shear to axions and to what extent can axion-feedback degeneracies be broken?
\item Can matter power spectrum reconstructions assist in distinguishing between axions and baryonic feedback processes, especially in the context of their ability to explain the $S_{8}$ tension?
\end{enumerate}

In Sec.~\ref{sec:modelling}, we describe the effects of axions on structure formation in more detail, as well as our approach to modelling axions and baryonic feedback together. In Sec.~\ref{sec:emulator}, we describe the training of an emulator to make this modelling approach possible in a Markov chain Monte Carlo (MCMC) analysis. In Sec.~\ref{sec:data}, we outline how we generate forecast LSST Y1 weak lensing data vectors. In Sec.~\ref{sec:results}, we present the forecast results and we conclude in Sec.~\ref{sec:conclusion}.

\section{Modelling axions and baryonic feedback}\label{sec:modelling}

The halo model for structure formation provides an analytic expression for the non-linear matter power spectrum $P_\mathrm{m}(k)$ \citep{Peacock_halomodel, Seljak_halomodel}. The model expresses $P_\mathrm{m}(k)$ as the sum of intra- and inter-halo components, assuming all dark matter exists in halos \citep{Press_Schechter}, i.e.
\begin{equation}\label{equ:halomodel}
    P_\mathrm{m}(k) = P^{1\rm{h}}(k) + P^{2\rm{h}}(k),
\end{equation}
where $P^{1\rm{h}}(k)$ is the intra- (or 1-) halo term and $P^{2\rm{h}}(k)$ is the inter- (or 2-) halo term. The 1-halo term is
\begin{equation}\label{equ:onehaloterm}
    P^{1\rm{h}}(k) = \int \text{d}M\,W^{2}(k,M,z)\,n(M,z)
\end{equation}
and the 2-halo term is
\begin{equation}\label{twohaloterm}
    P^{2\rm{h}}(k) = P_{\rm{L}}(k)\left[\int \text{d}M\,W(k,M,z)\,n(M,z)\,b(M,z) \right]^{2},
\end{equation}
where $n(M,z)$ is the halo mass function (HMF), $W(k,M,z)$ is the Fourier transform of the density profile \(\rho(r,M,z)\) of a halo of mass \(M\) at redshift \(z\) and radius \(r\), $b(M,z)$ is the halo bias and $P_{\rm{L}}(k)$ is the linear matter power spectrum.

\subsection{\LCDM halo model}\label{subsection:halomodel_LCDM}

In the standard \LCDM cosmological model, we assume the density profile of halos to follow the Navarro-Frenk-White (NFW) profile \citep{NFW}:

\begin{equation}\label{equ:NFWhalo}
    \rho(r,M) = \frac{\rho_{\rm{char}}}{r/r_{s}(1+r/r_{s})^{2}}
\end{equation}
with characteristic density $\rho_{\rm{char}}$ and scale radius $r_{\rm{s}}$. Thus,

\begin{equation}\label{equ:LCDM_NFW_WF}
    W(k,M,z) = \frac{1}{\bar{\rho}(z)}\int_{0}^{r_{\rm{v}}} \text{d}r \,4\pi r^{2} \frac{\sin(kr)}{kr} \rho(r,M,z),
\end{equation} 
where $\bar{\rho}(z)$ is the mean matter density of the Universe. We assume the halo does not extend beyond the virial radius $r_{\rm{v}}$. The halo mass is therefore
\begin{equation}\label{equ:halomass}
    M=\frac{4\pi}{3}\bar{\rho}(z)\,\Delta_{\rm{vir}}(z)\,r_{\rm{v}}^{3},
\end{equation}
where $\Delta_{\rm{vir}}(z)$ is the virial overdensity. The HMF \citep{Press_Schechter, Excursion_set}
\begin{equation}\label{equ:halomassfunction}
    n(M,z) = \frac{1}{M}\frac{\text{d}\Bar{n}}{\text{d}\ln{M}} = \frac{1}{2}\frac{\Bar{{\rho}}(z)}{M^{2}}f(\nu)\abs{\frac{\text{d}\ln \sigma^{2}(M,z)}{\text{d}\ln{M}}},
\end{equation}
where \(\Bar{n}\) is the halo number density, $\nu = \delta_{\rm{c}}(z)/\sigma(M,z)$, $\delta_{\rm{c}}$ is the critical value at which a spherical overdensity collapses and $\sigma(M,z)$ is the root mean square linear matter fluctuation in spheres that contain on average mass $M$. We use the multiplicity function $f(\nu)$ proposed by \citet{Sheth_Tormen}, which is a good fit to numerical simulations \citep[see e.g.][]{Reed_2003}. The halo bias is computed following \citet{Sheth_Tormen} \citep[for more discussion, see e.g.][]{vogt}.

Codes that implement the halo model have introduced a number of tweaks on top of the vanilla halo model presented above to match better to numerical simulations. These tweaks include a smoothing of the 1-halo to 2-halo transition, terms to account for halo bloating and scale dependence at small $k$, amongst other features. For a full discussion of the tweaks, see e.g. \citet{mead:2021} (Sec. 4 onwards). With these tweak factors, the halo model code \texttt{HMCODE} matches dark matter-only simulations to within \(\sim 5 \%\) for $k<10\,h\,\mathrm{Mpc}^{-1}$.

\subsection{Axion physics and ULA halo model}\label{sec:axionphysics}

The axion field $\phi$ obeys the Klein-Gordon equations of motion:
\begin{equation}\label{equ:KG_equation}
   \partial_{\mu}\partial^{\mu} \phi - m^{2}_{\rm{ax}}\phi = 0.
\end{equation}

\noindent The background component \(\phi_0 (t)\) obeys
\begin{equation}\label{equ:cosmoperturbations}
    \ddot{\phi}_{0} + 3H\dot{\phi}_{0} + m^{2}_{\rm{ax}}\phi_{0} = 0,
\end{equation}
where the Hubble parameter $H(t)=\Dot{a}/a$, \(a\) is the scale factor and dots denote derivatives with respect to cosmic time $t$. Due to Hubble friction, the axion field initially slow rolls when $H(t) \gg m_{\rm{ax}}$ and the axion energy density $\rho_{\rm{ax}}$ evolves like a cosmological constant.

At later times $t > t_{\rm{osc}}$, where ${H}(t_{\rm{osc}}) = m_{\rm{ax}}$, the axion field starts to oscillate and (in a time-averaged sense) $\rho_{\rm{ax}}$ evolves like pressureless matter. The axion is therefore CDM-like but with a scale-dependent sound speed \citep[for a full derivation, see e.g.][]{Marsh_axion}
\begin{equation}\label{equation:axionsoundspeed}
    c_{\rm{s}}^{2} = \frac{k^{2}/4m_{\rm{ax}}^{2}a^{2}}{1+ k^{2}/4m_{\rm{ax}}^{2}a^{2}}.
\end{equation}
At larger scales, this term tends to zero, matching axion behaviour to CDM. At smaller scales, the sound speed becomes non-negligible and thus these axion fluctuations oscillate, imprinting distinctive features into the CMB and large-scale structure. The Jeans wavenumber [Eq.~\eqref{equ:jeans_scale}] is the scale at which this transition from growth to oscillation occurs. For $k > k_{\rm{J}}$, the axion field suppresses growth of structure both because axion overdensities are oscillating and because this oscillation in turn suppresses growth of CDM overdensities. Boltzmann codes like \texttt{AxionCAMB} \citep{Hlozek_CMB_axions2015,axioncamb,Liu_axions} and \texttt{AxiCLASS} \citep{Axiclass_1,Axiclass_2} calculate these effects on the linear matter power spectrum and the CMB. In this work, we restrict ourselves to \(m_\mathrm{ax} \geq 10^{-27}\,\mathrm{eV}\), where \(t_\mathrm{osc} < t_\mathrm{eq}\) (\(t_\mathrm{eq}\) is the time of matter - radiation equality), where the axion starts behaving like matter before the matter-dominated epoch and the \(\Lambda\)-like behaviour at early times is not observationally relevant.

Using the ansatz \citep{Hwang_axion} $\phi = (2m_{\rm{ax}}a^3)^{-1/2}\left(\psi e^{im_{\rm{ax}}t} + \psi^{*} e^{-im_{\rm{ax}}t}\right)$, the full non-linear axion evolution is described by a coupled pair of Schr{\"o}dinger-Poisson equations:
\begin{subequations}
\begin{equation}\label{equation:poisson}
    i\dot{\psi} = - \frac{3i}{2}\frac{\dot{a}}{a} \psi -\frac{1}{2ma^2}\nabla^{2}\psi + m_{\rm{ax}}\Phi \psi,
\end{equation}
\begin{equation}\label{equation:schrodinger}
    \nabla^{2}\Phi = 4\pi G[m_{\rm{ax}}( |{\psi}|^{2}-\langle |{\psi}| \rangle^{2}) + \delta \rho],
\end{equation}
\end{subequations}
where angle brackets denote a spatial average, $\Phi$ is the gravitational potential and $\delta \rho$ is the perturbation to the CDM-baryon fluid. These equations are valid to arbitrarily small scales, capturing the wave-like behaviour of ULAs \citep{Widrow_schrodinger}. Studying these wave effects in simulations, e.g., \citet{Schive_axiondensity,Schwabe_sims, May_Springel_FDM, Zimmerman_FDM} highlight that axions behave like CDM on large scales (consistent with Eq.~\eqref{equation:axionsoundspeed}), but, at scales below the de Broglie length, form solitons, interference fringes and dark matter granules.

\texttt{axionHMCODE} \citep{vogt} presents a halo model approach for capturing the effect of axions on the non-linear matter power spectrum, as an extension to the \LCDM halo model approach presented in \texttt{HMCODE} \citep[][]{mead:2021}. We use this halo model rather than running hydrodynamical mixed DM simulations due to the computational expense of solving the axion Schr{\"o}dinger-Poisson equations. These simulations are expensive owing to the number of spatial elements and the time-step requirements needed to resolve the oscillatory features of FDM from the de Broglie wavelength to cosmological scales, especially alongside the effects of baryonic feedback \citep{Mocz_HYDROAXION}. Further, weak lensing is less sensitive to the one-halo term where explicitly fuzzy phenomena manifest \citep{May_Springel_FDM,Rogers_S8tensionaxions,Lague}, thus eliminating the need for modelling accurately with simulations to such small scales.

The perturbation to the total matter density \(\delta_\mathrm{m}\) can be expressed as a weighted sum of the perturbations to the combined CDM and baryon \(\delta_\mathrm{c}\) and axion \(\delta_\mathrm{ax}\) components:
\begin{equation}\label{equ:mixedoverdensities}
    \delta_{\rm{m}} = \frac{\Omega_{\rm{c}}}{\Omega_{\rm{m}}}\delta_{\rm{c}} + \frac{\Omega_{\rm{ax}}}{\Omega_{\rm{m}}}\delta_{\rm{ax}},
\end{equation}
where $\Omega_{\rm{c}}$ is the combined CDM and baryon energy density and $\Omega_{\rm{ax}}$ is the axion energy density. Thus, the total matter power spectrum
\begin{equation}
\label{eq:power_cold_axion_no_feedback}
    P_{\rm{m}}(k) = \left(\frac{\Omega_{\rm{c}}}{\Omega_{\rm{m}}}\right)^{2}P_{\rm{c}}(k) +  \frac{2\Omega_{\rm{c}}\Omega_{\rm{ax}}}{\Omega_{\rm{m}}^2}P_{\rm{c,ax}}(k)   + \left(\frac{\Omega_{\rm{ax}}}{\Omega_{\rm{m}}}\right)^{2}P_{\rm{ax}}(k),
\end{equation}
where $P_{\rm{c}}(k)$ is the combined CDM and baryon power spectrum, $P_{\rm{ax}}(k)$ is the axion power spectrum and $P_{\rm{c,ax}}(k)$ is the power spectrum crossed between the cold and axion components. Here and onwards, we refer to the combined CDM and baryon component simply as the cold component since we do not consider feedback effects until Sec.~\ref{sec:feedback}. Since we are in the regime where the cold component is larger than the axion component (\(m_\mathrm{ax} \sim 10^{-25}\,\mathrm{eV}\)), the main power contribution comes from CDM, but oscillatory features are imprinted on $P_{\rm{m}}(k)$ through the axion and cross-power components.

The cold component \(P_\mathrm{c}(k)\) follows the halo model given in Sec.~\ref{subsection:halomodel_LCDM}, as we still assume all the CDM and baryons are bound into halos, i.e. Eq.~\eqref{equ:halomodel} with the NFW halo density profile [Eq.~\eqref{equ:NFWhalo}] and Press-Schechter HMF [Eq.~\eqref{equ:halomassfunction}]. However, the axion perturbation is treated differently:
\begin{equation}
    \delta_{\rm{ax}} = F_{\rm{h}}\delta_{\rm{h}} + (1-F_{\rm{h}})\delta_{\rm{L}},
\end{equation}
where $\delta_{\rm{h}}$ and $\delta_{\rm{L}}$ are the halo and linear perturbations respectively, i.e. the axion fluctuation is split into components that respectively cluster into halos and do not cluster into halos, where $F_{\rm{h}}$ is the ULA fraction in halos. This heuristic approach by \citet{Marsh_Silk} posits that no axion halo will form if the Jeans scale is larger than the virial radius. \citet{vogt} use this approach to infer a cut-off halo mass below which axions do not cluster into halos. This approach is similar to the halo model treatment of neutrinos in \citet{mead:2021}, which separates neutrinos into components that do and do not cluster into halos. The cross and axion power spectra are thus respectively:
\begin{subequations}
\begin{eqnarray}
    P_{\rm{c,ax}}(k) &=& F_{\rm{h}} P^{\rm{h}}_{\rm{c,ax}}(k) + (1-F_{\rm{h}})\sqrt{ P_{\rm{c}}(k) P^{\rm{L}}_{\rm{ax}}(k)},  \hspace{0.3 truein}  \\
    P_{\rm{ax}}(k) &=& F^{2}_{\rm{h}} P^{\rm{h}}_{\rm{ax}}(k) + 2(1-F_{\rm{h}})\sqrt{ P^{\rm{h}}_{\rm{ax}}(k) P^{\rm{L}}_{\rm{ax}}(k)} \nonumber  \\
  & &  \hspace{0.6 truein} +(1-F_{\rm{h}})^{2}P^{\rm{L}}_{\rm{ax}}(k).
\end{eqnarray}
\end{subequations}

The linear axion power spectrum $P^{\rm{L}}_{\rm{ax}}(k)$ is computed by \texttt{AxionCAMB} using axion transfer functions of the primordial power spectrum. The non-linear halo components $P^{\rm{h}}_{\rm{ax}}(k)$ and $P^{\rm{h}}_{\rm{c, ax}}(k)$ are computed using the halo model expansion of Eq.~\eqref{equ:halomodel} as described in detail by \citet{vogt}. There are two main physical differences from the \(\Lambda\)CDM halo model in Sec.~\ref{subsection:halomodel_LCDM}. First, the NFW density profile in Eq.~\eqref{equ:NFWhalo} is replaced by a hybrid profile, which is NFW-like in the outskirts of the axion halo but has a soliton core (rather than cusp) in the central region. The core density profile for the soliton component is a fit to the simulations of \citet{Schive_axiondensity}:
\begin{equation}\label{equ:soliton_core}
    \rho_{\mathrm{c}}(r) = \frac{1.9(1+z)}{(1+0.091(r/r_{\mathrm{c}})^{2})^{8}}\times \left( \frac{r_{\mathrm{c}}}{\mathrm{kpc}} \right)^{-4}  \left( \frac{m_{\mathrm{ax}}}{10^{-23}\mathrm{eV} } \right)^{-2},
\end{equation}
in units of $M_{\odot}\,\mathrm{pc}^{-3}$, where the soliton core radius \(r_\mathrm{c}\) is as defined by \citet{vogt}.

Second, the halo mass function in Eq.~\eqref{equ:halomassfunction} changes for both cold and axion halos. We follow a biased tracer approach where every cold halo hosts an axion halo, as long as the axion halo mass is greater than the cut-off detailed above. This means that $n(M_\mathrm{c})\mathrm{d}M_{\mathrm{c}} = n(M_\mathrm{ax})\mathrm{d}M_{\mathrm{ax}}$, where \(M_\mathrm{c}\) and \(M_\mathrm{ax}\) are respectively the cold and axion halo masses. We then define the mass \(M\) in Eq.~\eqref{equ:halomassfunction} to be the sum of the cold and axion halo masses. We also follow the conclusions found in \(N\)-body simulations with massive neutrinos, which find a better fit to the HMF of Eq.~\eqref{equ:halomassfunction} if only the cold overdensity $\delta_{\mathrm{c}}$ is used in its calculation \citep{Villaescusa-Navarro_3,Villaescusa-Navarro_2,Villaescusa-Navarro_1}. This approach again follows \citet{vogt}, which treats ULAs as a biased tracer of CDM and baryons, similar to the halo model approach used for neutral hydrogen \citep{Padmanabhan_tracer} and neutrinos \citep{Massara2024} .

\begin{figure*}

    \begin{minipage}{0.33\textwidth}
    	\includegraphics[width=1.\columnwidth]{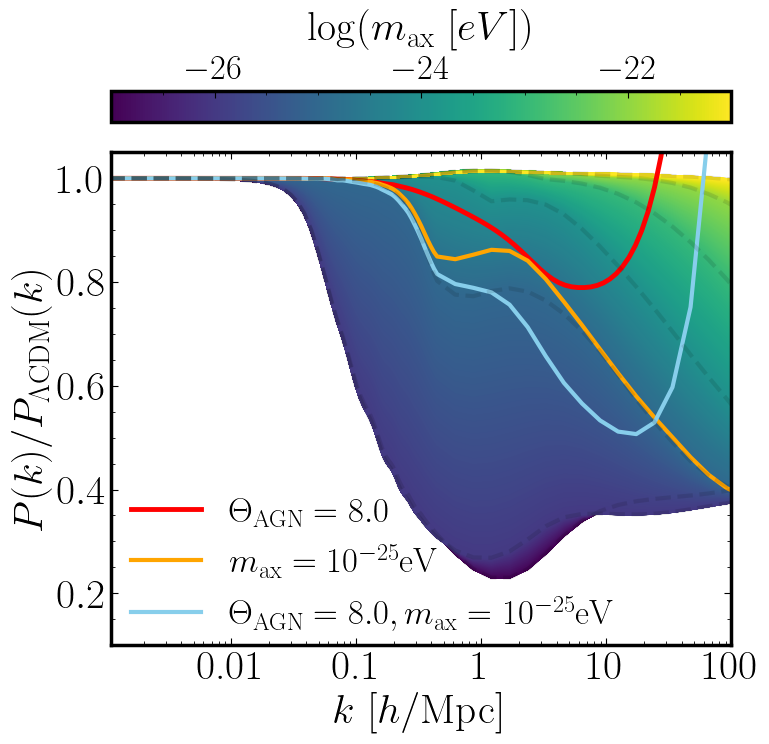} 
    \end{minipage}
    \begin{minipage}{0.33\textwidth}
        \includegraphics[width=1.\columnwidth]{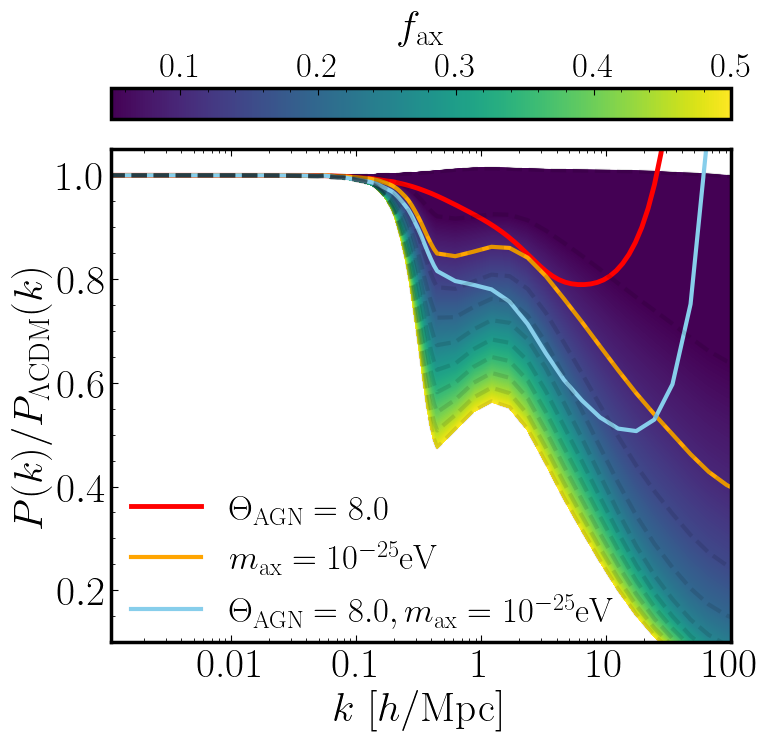} 
    \end{minipage}
    \begin{minipage}{0.33\textwidth}
        \includegraphics[width=1.\columnwidth]{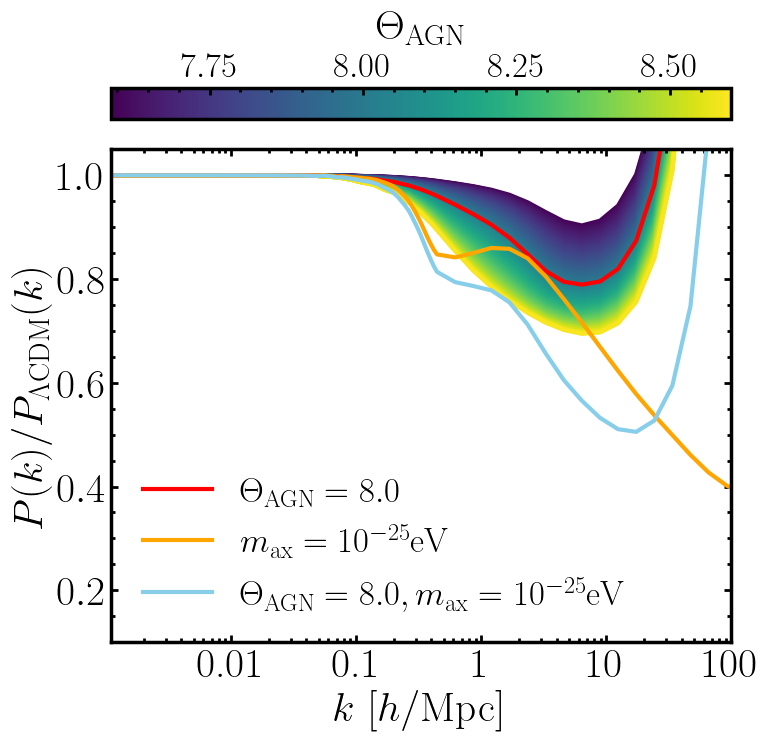} 
    \end{minipage}
\caption{\textit{Left}: the effect of varying axion mass \(m_\mathrm{ax}\) on the matter power spectrum \(P_\mathrm{m} (k)\), where \(m_\mathrm{ax} = 10^{-25}\,\mathrm{eV}\) is highlighted in orange. The axion fraction \(f_\mathrm{ax}\) is fixed to 0.1 and feedback is not included, i.e. the model in Eq.~\eqref{eq:power_cold_axion_no_feedback}. \textit{Centre}: the effect of varying axion fraction \(f_\mathrm{ax}\) on \(P_\mathrm{m} (k)\), where \(f_\mathrm{ax} = 0.1\) is highlighted in orange. The axion mass \(m_\mathrm{ax}\) is fixed to $10^{-25}\,\mathrm{eV}$ and feedback is not included, i.e. the model in Eq.~\eqref{eq:power_cold_axion_no_feedback}. \textit{Right}: the effect of varying feedback parameter \(\Theta_\mathrm{AGN}\) on \(P_\mathrm{m} (k)\), where \(\Theta_\mathrm{AGN} = 8.0\) is highlighted in red. We use the \(\Lambda\)CDM halo model without axions but with feedback, i.e. as presented in Sec.~\ref{subsection:halomodel_LCDM} but with the window function in Eq.~\eqref{equ:LCDMbaryonWFC}. In each panel, the same combined axion and feedback model, i.e. Eq.~\eqref{equ:total_pk}, is highlighted in blue and we always show the ratio to the \(\Lambda\)CDM, feedback-free limit \(P_{\Lambda\mathrm{CDM}}(k)\). A lighter axion mass generally suppresses the matter power spectrum to increasingly large scales and a larger fraction suppresses more at a given wavenumber, introducing more complex shapes in power suppression compared to feedback.}

\label{fig:Pk_suppression_plots}

\end{figure*}

We assume that this mixed axion and CDM halo model is a reasonable approximation for $m_{\rm{ax}} \geq 10^{-27}\,\mathrm{eV}$ (when ULAs are matter-like since before matter - radiation equality) forming a fraction $f_\mathrm{ax}<50 \%$ of the total DM. The above formalism does not include the effect of baryonic feedback on the matter power spectrum.\footnote{Tweaks to the axion halo model presented in \citet{Dome_Tibor} are not included as they are not relevant to the redshifts we consider (\(z < 1\)), nor are they consistent with the baryonic feedback correction that we will introduce in Sec. \ref{sec:feedback}.}

\subsection{Baryonic feedback in the mixed cold and axion halo model}
\label{sec:feedback}
Baryonic feedback is expected to affect the small-scale matter power spectrum \citep[see e.g.][and references therein]{Chisari_2019, vandaalen:2020}. The exact shape and amplitude of this effect are still uncertain and vary between different hydrodynamical simulations with sensitivity to resolution, box size and sub-grid prescriptions for implementing feedback processes \citep[see e.g.][]{BAHAMAS,flamingo}. The exact effect also depends on which data are used to calibrate each model. X-ray measurements of gas fractions in groups and clusters \citep{Schneider:2020}, weak lensing and Sunyaev-Zeldovich measurements each prefer different feedback prescriptions \citep{CPAAGPE, Bigwood:2024}.

The halo model approach of \texttt{HMCODE} \citep{mead:2021} has an extension to include the effects of feedback in a cosmological model with \(\Lambda\)CDM and massive neutrinos, with parameters calibrated to reproduce matter power spectra measured in the BAHAMAS \citep{BAHAMAS} and COSMO-OWLS \citep{COSMOowls} suites of hydrodynamical simulations. \citet{mead:2021} replace the window function \(W(k,M,z)\) of Eq.~\eqref{equ:LCDM_NFW_WF} with
\begin{equation}\label{equ:LCDMbaryonWFC}
    \Tilde{W}(k,M,z) = \left(\frac{\Omega_{\rm{CDM}}}{\Omega_{\rm{CDM}}+\Omega_{\rm{b}}} + f_{\rm{g}}(M)\right)W(k,M,z) + f_{*}\frac{M}{\bar
    {\rho}},
\end{equation}
where \(\Omega_\mathrm{CDM}\) is the CDM energy density, \(\Omega_\mathrm{b}\) is the baryon energy density,
\begin{equation}
     f_{\rm{g}} (M)= \left(\frac{\Omega_{\rm{b}}}{\Omega_{\rm{CDM}}+\Omega_{\rm{b}}} - f_{*}\right)\frac{(M/M_{\rm{b}})^{\beta}}{1+(M/M_{\rm{b}})^{\beta}}
\end{equation}
and $\beta$, $f_{*}$ and $M_{\rm{b}}$ are parameters fit to simulations. These parameters are related to a single feedback strength parameter $\Theta_{\rm{AGN}}$ (see Sec. 6.3 of \citet{mead:2021} for more details).\footnote{This feedback model also has a percent-level dependence on cosmology, specifically $\Omega_{\mathrm{m}}$ and $\Omega_{\mathrm{b}}$, despite being calibrated to a few simulations with only best fit WMAP- and \textit{Planck}-like cosmologies.} There are alternative models based on emulators, e.g. \texttt{BACCO} and \texttt{BCemu} \citep{Bacco, BCEmu}. These models have several feedback parameters, calibrated to reproduce a range of different simulations, rather than the BAHAMAS and COSMO-OWLS simulations. For this work, we restrict ourselves to considering the one-parameter $\Theta_{\rm{AGN}}$ model from \texttt{HMCODE}, but future surveys may implement these more flexible baryonic feedback models.

We make the ansatz that a universe containing axions, CDM and baryonic feedback will have a non-linear matter power spectrum
\begin{eqnarray}
    P_{\rm{m}} \hspace{-2.5mm} & = & \hspace{-2.5 mm} P_{\texttt{EucEmu}}^{\rm{\Lambda CDM}}(\mathbf{\theta}_{\Lambda\mathrm{CDM}}) \hspace{-1mm} \ \times \ \hspace{-1mm} \frac{P_{\rm{\texttt{axHMCODE}}}^{\rm{axions}} (m_\mathrm{ax}, f_\mathrm{ax},\mathbf{\theta}_{\Lambda\mathrm{CDM}})}{P_{\rm{\texttt{axHMCODE}}}^{\rm{\Lambda CDM}} (f_\mathrm{ax} = 0, \mathbf{\theta}_{\Lambda\mathrm{CDM}})} \ \ \  \nonumber \\
    & & \times \ \frac{P_{\rm{\texttt{HMCODE}}}^{\rm{feedback}}(\Theta_\mathrm{AGN}, \mathbf{\theta}_{\Lambda\mathrm{CDM}})}{P_{\rm{\texttt{HMCODE}}}^{\rm{no-feedback}} (\mathbf{\theta}_{\Lambda\mathrm{CDM}})},  \label{equ:total_pk}
\end{eqnarray}
where \(\mathbf{\theta}_{\Lambda\mathrm{CDM}}\) is a vector of \(\Lambda\)CDM parameters, the first term is the matter power spectrum from \texttt{EuclidEmulator} \citep{EuclidEmulator}, the second is the non-linear correction due to axions predicted by \texttt{axionHMCODE} (Eq.~\eqref{eq:power_cold_axion_no_feedback}) and the third is the non-linear correction due to baryonic feedback predicted by the halo model extension from \texttt{HMCODE} (Eq.~\eqref{equ:halomodel} using Eq.~\eqref{equ:LCDMbaryonWFC}). We use \texttt{EuclidEmulator} as the baseline feedback-free \(\Lambda\)CDM matter power spectrum as this most closely matches the results of numerical simulations (more so than the \(\Lambda\)CDM halo model introduced in Sec. \ref{subsection:halomodel_LCDM}).

In combining the feedback-affected CDM and baryon halo model (third term) with the halo model for ULA cosmologies, we are effectively assuming that baryonic feedback affects all components as if they were CDM (in particular, that \(\Theta_\mathrm{AGN}\) is uncorrelated with \(m_\mathrm{ax}\) and \(f_\mathrm{ax}\)). There are currently no cosmological hydrodynamical simulations of mixed cold and axion DM interacting with baryons that span the axion mass, halo mass and redshift ranges that we consider here and therefore no explicit numerical checks of the model. Nonetheless, this feedback model has been shown to reproduce simulations with varying neutrino masses, the effect of which is qualitatively similar to axions (a fraction not clustering into halos leading to small-scale power suppression). We therefore anticipate that this a good approximation, at least as accurate an assumption as for \(\Lambda\)CDM cosmologies with massive neutrinos. Our analysis is a first step towards assessing the interplay between baryonic feedback and axions and we advocate for more detailed studies with simulations in the future.

Figure \ref{fig:Pk_suppression_plots} shows the effects of varying axion mass, axion fraction and feedback strength on the power spectrum model in Eq.~\eqref{equ:total_pk}. In each panel, the same axion-only, feedback-only and combined models are shown for comparison. The effect of decreasing axion mass at fixed axion fraction is to shift the suppression to larger scales. This shift arises largely because, as the mass decreases, the axion Jeans wavenumber decreases. Increasing axion fraction at fixed axion mass increases the amplitude of suppression whilst weakly affecting the suppression scale. There is a regime at $m_{\mathrm{ax}} \sim 10^{-21}\,\mathrm{eV}$ and $f_{\mathrm{ax}}\sim10\%$ where there is a small boost in power at $k\sim1\,h\,\rm{Mpc}^{-1}$ compared to the \LCDM baseline. This boost is the result of the coherence of the axion soliton coinciding with the 1-halo to 2-halo transition regime \citep{vogt}. The baryonic feedback model has two characteristic features: an initial suppression resulting from gas expulsion smoothing out the matter distribution and then a high-\(k\) boost from galaxy formation at halo centres. Increasing feedback strength pushes the characteristic bell-shape suppression to higher \(k\) and greater amplitude. We note that this is only one model of feedback and there is in general variety in the predictions from different hydrodynamical simulations.

\section{Emulating axions}\label{sec:emulator}

Solving the equations governing linear and non-linear axion evolution (Sec.~\ref{sec:axionphysics}) is time intensive. One evaluation of \texttt{axionHMCODE} can take up to 60 seconds to produce a power spectrum at one redshift. It is thus computationally demanding to sample numerically posterior distributions, where we will need \(\mathcal{O}(10^6)\) code evaluations. We therefore construct an emulator of \texttt{axionHMCODE} that will evaluate the numerator and denominator of the second term in Eq.~\eqref{equ:total_pk}. An emulator is a machine learning model that learns the mapping from parameters to observables, trained on a set of code evaluations like \texttt{axionHMCODE} \citep{Heitmann:2009cu,Rogers:2018smb,cosmopower}. Once trained, it massively speeds up the generation of observables like power spectra.

For the training, validation and test sets, we generate 200,000 non-linear matter power spectra using \texttt{axionHMCODE}, sampling the prior volume defined in Table \ref{tab:training_data_grid} using a Latin hypercube. This parameter volume matches that used by \texttt{EuclidEmulator} \citep{EuclidEmulator} to ensure consistency in Eq.~\eqref{equ:total_pk}, where the first term is given by \texttt{EuclidEmulator}. Each power spectrum has 400 samples equally spaced in $\log (k)$ in the interval \(k \in [10^{-3}, 10^2]\,h\,\mathrm{Mpc}^{-1}\). We use $80\%$ of our data as a training/validation set for the emulator, with the remaining $20\%$ used as a test data set. For the emulator, we use the neural architecture presented in \texttt{axionEmu}\footnote{\url{https://github.com/keirkwame/axionEmu}.}, itself based on a modified version of \texttt{CosmoPower} \citep{cosmopower}. The neural network is a sequence of four hidden layers (each with 512 nodes) that map eight parameters (Table \ref{tab:training_data_grid}) to the non-linear matter power spectrum evaluated at 400 \(k\) values. We use the custom activation function introduced in \citet{cosmopower}. The initial learning rate is $10^{-2}$, decreasing in powers of ten in each epoch to $10^{-6}$.

\begin{table}
    \caption{Parameter volume over which the \texttt{axionHMCODE} emulator is trained.}
\label{tab:training_data_grid}
\begin{center}
\begin{tabular}{lll}
Parameter & Emulator range \tabularnewline
\hline 
Total matter density $\Omega_{\rm m}$ & $[0.25, 0.4]$   \tabularnewline
Baryon density $\Omega_{\rm b}$ & $[0.03, 0.07]$  \tabularnewline
Primordial scalar amplitude $10^{-9}A_{\rm s}$ &  $[0.5, 5.0]$ \tabularnewline
Primordial scalar spectral index $n_{\rm s}$ & $[0.87,1.07]$  \tabularnewline
Dimensionless Hubble parameter $h$ & $[0.5, 0.8]$  \tabularnewline
log(Axion mass $[{\rm eV}]$) $\log\left({m_{\rm{ax}}\,[\mathrm{eV}]}\right)$ & $[-28,-20]$  \tabularnewline
Axion fraction $f_{\rm{ax}}$ & $[0,1]$
\tabularnewline
Redshift $z$ & $[0,5]$
\tabularnewline
\hline 

\end{tabular}
\end{center}
\end{table}

\begin{figure}
\centering
    \includegraphics[width=1\columnwidth]{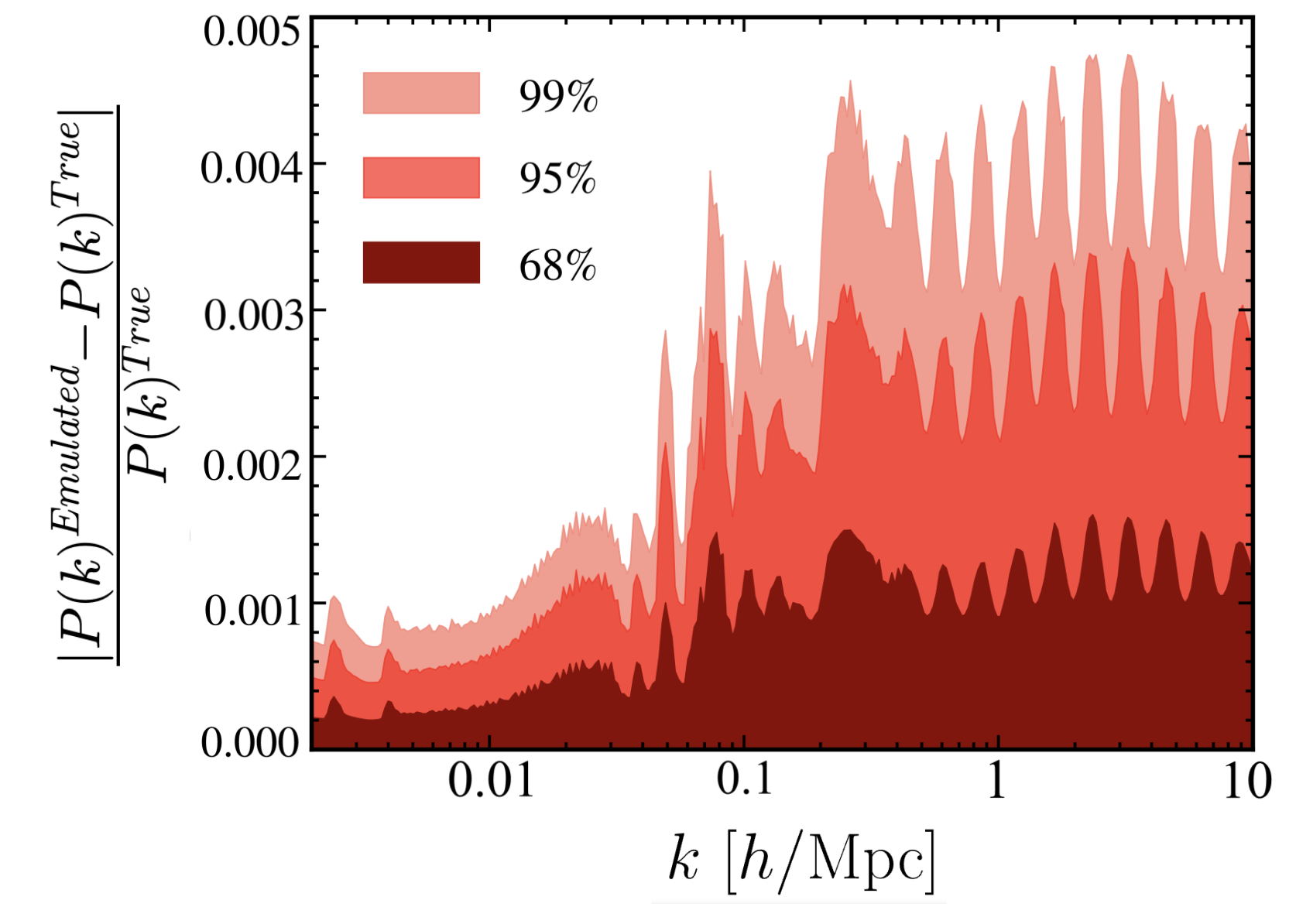} 
\caption{Accuracy of the emulator for \texttt{axionHMCODE} mixed cold and axion non-linear power spectra. We show the 68\%, 95\% and 99\% upper limits on the distribution of the absolute difference between emulator prediction and test power spectrum in ratio to the test value. We achieve sub-percent accuracy up to $k=10\,h\,\rm{Mpc}^{-1}$.}
\label{fig:emulator_accuracy}
\end{figure}

The accuracy of the emulator compared to test power spectra from \texttt{axionHMCODE} as a function of $k$ is shown in Fig.~\ref{fig:emulator_accuracy}. Errors grow on smaller scales as oscillatory features in the axion power spectra become more prominent and are harder to emulate. Errors are smaller on larger scales as ULAs behave like CDM on scales larger then the Jeans length, which is easier to emulate. Errors remain at the sub-percent level in $99\%$ of cases for $k\leq10~h\,\rm{Mpc}^{-1}$. This is sufficient as the data error for LSST Y1 is much larger than this error on the emulator \citep[see e.g.][]{Preston_2024}. We also test specifically the accuracy of the emulator at the edges of the Latin hypercube, ensuring there are no divergent errors in this regime. For a real data anaylsis, it will be important to quantify the errors on the predictions from \texttt{axionHMCODE}, not just the errors on the emulator of the halo model. Comparisons of simulations against predictions from halo models can be incorporated into the covariance \citep{Koyama_Euclid}.

\section{Forecast
cosmic shear data from LSST Year 1}\label{sec:data}

\begin{table}
    \caption{The \(\Lambda\)CDM, axion, feedback and systematics parameters that we vary in the analysis (\textit{left}), quoted with their assumed prior distributions (\textit{centre}) and the input ``truth'' values that we use when generating fake LSST Y1-like shear data (\textit{right}). To model galaxy intrinsic alignments 
    (IA), we use the redshift-dependent non-linear alignment (NLA) model and the LSST Y1-like calibration priors are derived from \citet{DESC_requirements}.}
\label{tab:priors}
\begin{center}
\begin{tabular}{lll}
Parameter & Prior & Truth \tabularnewline
\hline 
\bf{\(\Lambda\)CDM} \tabularnewline
Total matter density $\Omega_{\rm m}$ & $\mathcal{U}(0.25, 0.4)$  & 0.3135 \tabularnewline
Baryon density $\Omega_{\rm b}$ & $\mathcal{U}(0.03, 0.07)$  & 0.0489 \tabularnewline
Primordial amplitude $10^{-9}A_{\rm s}$ &  $\mathcal{U}(0.5, 5.0)$ & 2.091 \tabularnewline
Primordial spectral index $n_{\rm s}$ & $\mathcal{U}(0.87,1.07)$  & 0.9671 \tabularnewline
Hubble parameter $h$ & $\mathcal{U}(0.5, 0.8)$ & 0.6744 \tabularnewline
\hline
\bf{Extensions} \tabularnewline
log(Axion mass $[{\rm eV}]$) $\log({m_{\rm{ax}}\,[\mathrm{eV}]})$ & $\mathcal{U}(-28,-20) $  & Varies \tabularnewline
Axion fraction $f_{\rm{ax}}$ & $\mathcal{U}(0,1) $ & Varies \tabularnewline
Feedback strength $\Theta_{\rm{AGN}}$ & $\mathcal{U}(7.3,9.3)$  & Varies \tabularnewline
\hline 
\bf{Systematics} \tabularnewline
IA NLA amplitude $A$ & $\mathcal{U}(-0.6,1.2)$  & 0.0 \tabularnewline
IA NLA redshift index $\eta$ & $\mathcal{U}(0,5)$ & 0.0   \tabularnewline
Source redshift bin 1 $\Delta z^1$ & $\mathcal{N}(0, 0.002491)$  & 0.0 \tabularnewline
Source redshift bin 2 $\Delta z^2$ & $\mathcal{N}(0, 0.002903)$  & 0.0 \tabularnewline
Source redshift bin 3 $\Delta z^3$ & $\mathcal{N}(0, 0.003302)$  & 0.0 \tabularnewline
Source redshift bin 4 $\Delta z^4$ & $\mathcal{N}(0, 0.003824)$  & 0.0 \tabularnewline
Source redshift bin 5 $\Delta z^5$ & $\mathcal{N}(0, 0.005062)$ & 0.0 \tabularnewline
Shear calibration 1 $m^1$ & $\mathcal{N}(0, 0.013)$ & 0.0 \tabularnewline
Shear calibration 2 $m^2$ & $\mathcal{N}(0, 0.013)$ & 0.0 \tabularnewline
Shear calibration 3 $m^3$ & $\mathcal{N}(0, 0.013)$ & 0.0 \tabularnewline
Shear calibration 4 $m^4$ & $\mathcal{N}(0, 0.013)$ & 0.0 \tabularnewline
Shear calibration 5 $m^5$ & $\mathcal{N}(0, 0.013)$ & 0.0 \tabularnewline
\hline 
\end{tabular}
\end{center}
\end{table}

In the small-angle (Limber) approximation, the shear correlation function between redshift bins $i$ and $j$ is given by
\begin{equation}
\xi_{+/-}^{ij}(\theta) = \frac{1}{2\pi} \int \mathrm{d} \ell\,\ell\,J_{0/4}(\ell\theta)\,C_{\kappa}^{ij}(\ell),
\label{equ:totalangspectra}
\end{equation}
where \(\ell\) is multipole and the convergence angular power spectrum \(C_\kappa^{ij}(\ell)\) is related to the matter power spectrum $P_{\rm{m}}$ by
\begin{equation}
C_\kappa^{ij} (\ell) = \int_0^{\chi_{\rm H}} \mathrm{d}\chi {q_i(\chi) q_j(\chi) \over f_K^2(\chi) }  P_{\rm m} \left(\frac{\ell}{f_K(\chi)}, \chi\right). \label{equ:Ckappa1}
\end{equation}
In Eq.~\eqref{equ:Ckappa1}, $\chi$ is comoving radial distance, \(\chi_\mathrm{H}\) is the comoving Hubble radius and $f_K(\chi)$ is the comoving angular diameter distance at $\chi$ in a Friedmann-Lema\^{i}tre-Robertson-Walker background with curvature parameter $K$. The lensing efficiency functions are given by
\begin{equation}
    q_i(\chi) = {3H_0^2 \Omega_{\rm m} \over 2c^2} {f_K(\chi) \over a(\chi)} 
    \int_\chi^{\chi_\mathrm{H}} \mathrm{d}\chi^\prime n_i\left(\chi^\prime\right) {f_K (\chi^\prime - \chi) \over f_K(\chi^\prime)},   \label{equ:Ckappa2}
\end{equation}
where $a(\chi)$ is the dimensionless scale factor and $n_i(\chi)$ is the galaxy redshift distribution in tomographic bin $i$ normalised so that $\int \mathrm{d}\chi\,n_i(\chi) = 1$  \citep[see e.g.][]{Hildebrandt:2017}. In this work, we assume zero curvature\footnote{This assumption is consistent with the strong observational constraint that the curvature energy density today $\Omega_{K} = 0.0004 \pm 0.0018$ \citep{Efstathiou:2020}.} so $f_K(\chi) = \chi$. The effects of axions and feedback that we investigate here enter through \(P_\mathrm{m}(k,z)\) in Eq.~\eqref{equ:Ckappa1} as modelled in Eq.~\eqref{equ:total_pk}.

We generate cosmic shear correlation functions with forecast \textit{Rubin} LSST Y1 survey properties \citep{DESC_requirements}. The LSST Y1 photometric galaxy survey is forecast to have five tomographic source redshift bins covering $0<z<2$ \citep[for more discussion on the forecast LSST redshift distribution, see e.g.][]{Paopiamsap_DESC_IA}. We fix the expected number density of galaxies $n_{\rm{{gal}}} = 10\,\mathrm{arcmin}^{-2}$ and the root mean square of the typical intrinsic galaxy ellipticity $\sigma_{e} = 0.26$ \citep{Heymans:2013}. The expected survey footprint covers $18000~\rm{deg}^{2}$ of the southern hemisphere. We use these specifications to generate fake LSST Y1 shear correlation functions for redshift bin pairs (see Appendix \ref{sec:app_A} for example sets of correlation functions).

When generating fake data, we fix the five \(\Lambda\)CDM parameters to the best-fit values given \Planck temperature and polarisation data \citep[][]{Params:2018} and vary \(m_\mathrm{ax}\), \(f_\mathrm{ax}\) and \(\Theta_\mathrm{AGN}\) according to the setting. We make our fake data free of systematic errors. The cosmological, astrophysical and systematics parameters and their priors are summarised in Table \ref{tab:priors}. When sampling the forecast posterior distributions, we marginalise over \(\Lambda\)CDM, axion and feedback parameters according to the setting. We always marginalise over uncertainties in the central redshift of each tomographic bin $\Delta z^{i}$ and galaxy shape calibration uncertainties in each bin $m^{i}$, alongside parameters of the non-linear alignment (NLA) intrinsic galaxy alignment model \citep{Bridle_KingNLA}. The NLA model uses the growth factor $D(z)$, which in this work accounts for the presence of axions, as well as $\Omega_{\mathrm{m}}$ which is the total matter density of baryons, CDM and axions \citep[for more details on IA formalisms, see the review by e.g.][]{Lamman:2024}. \citet{Preston_2024} highlight the insensitivity of constraints from weak lensing to different true values of the NLA amplitude and redshift index, thus we fix the true values in this work. Next-generation surveys are likely to include sample selection to mitigate for the effects of IA, such as constructing a blue-only shear sample \citep{McCullough_BlueIA, Siegel_DESI_IA}. We use the \texttt{CosmoSIS} inference package \citep{Zuntz:2015} to sample forecast posterior distributions using the \texttt{MultiNest} nested sampling algorithm \citep{Feroz_2009}, using a stopping tolerance $\epsilon=0.01$.

The strength of feedback is currently uncertain; for more discussion on modelling baryonic feedback, see e.g. \citet{Bigwood:2024} and references therein. Given the current uncertainty in the effects of feedback on the matter power spectrum, it is important to marginalise with a flexible feedback model in order to capture its effect. To this end, we use an extended uniform prior in Table \ref{tab:priors} on $\Theta_{\rm{AGN}}$ with the range \([7.3, 9.3]\) compared to the standard range of \([7.6, 8.0]\) \citep{vanDaalen:2011,BAHAMAS}, the latter of which reflects the range of simulations against which the feedback model is calibrated \citep{mead:2021,Bigwood:2024}.

\section{Results}\label{sec:results}

\subsection{The effects of ignoring axions if they exist}
\label{sec:ignore_axions}

\begin{figure*}
    \begin{minipage}{0.495\textwidth}
    \includegraphics[width=1.\columnwidth]{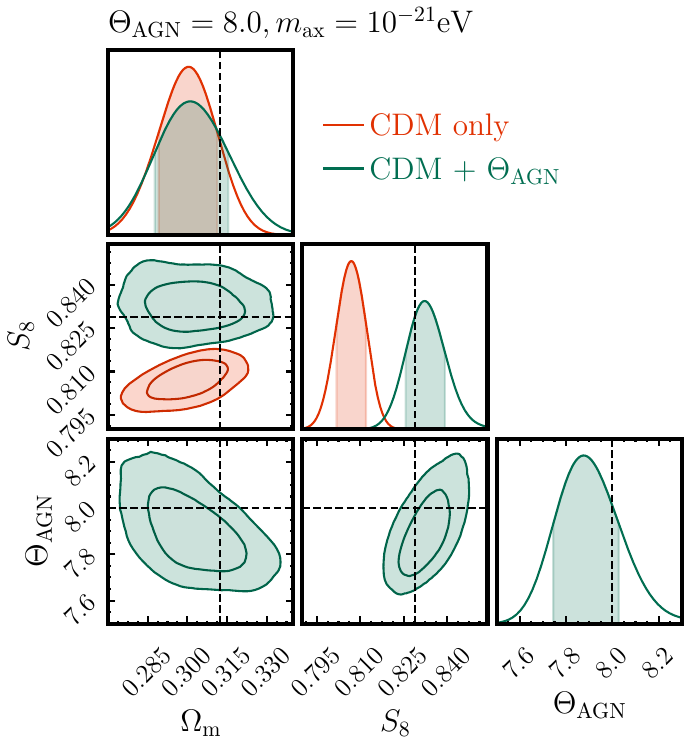} 
    \end{minipage}
    \begin{minipage}{0.495\textwidth}
        \includegraphics[width=1.\columnwidth]{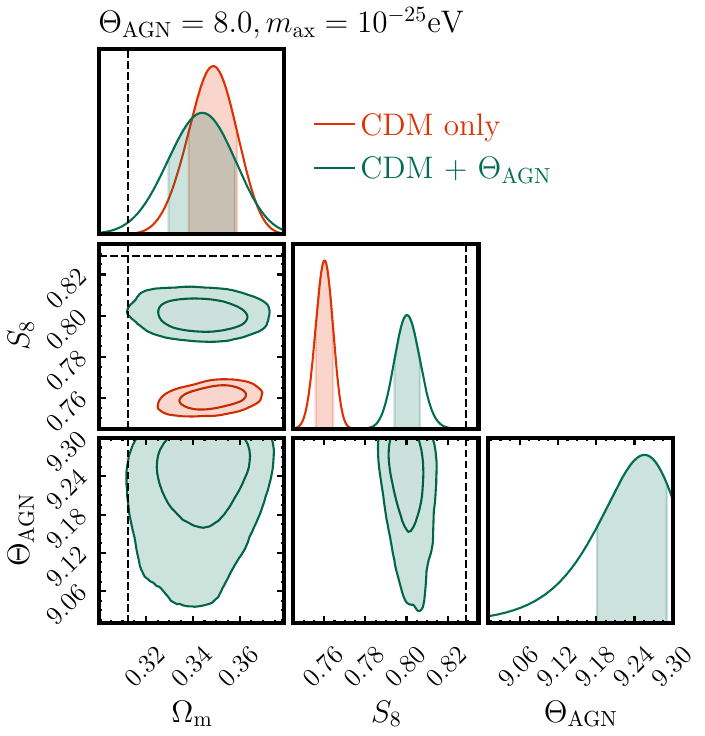} 
    \end{minipage}

\caption{Posterior distributions from a \LCDM analysis of a simulated LSST Y1-like shear correlation function that includes the effects of both axions and feedback. One analysis accounts for baryon feedback by varying the \(\Theta_\mathrm{AGN}\) parameter (green), while the other neglects to account for this effect (red) and neither analysis models the axion dark matter component. \textit{Left}: the fake data are generated for $m_{\rm{ax}}=10^{-21}\,\mathrm{eV}$, \(f_\mathrm{ax} = 0.1\) and $\Theta_{\rm{AGN}}=8.0$; \textit{right}: $m_{\rm{ax}}=10^{-25}\,\mathrm{eV}$, \(f_\mathrm{ax} = 0.1\) and $\Theta_{\rm{AGN}}=8.0$. We report marginalised posteriors of $\Omega_{\rm{m}}$, $S_{8}$ and $\Theta_{\rm{AGN}}$; the inner and outer contours respectively indicate the 68\% and 95\% credible regions. The true parameter values are shown as dashed black lines. We note that the true $S_{8}$ includes the effect of axions on the linear matter power spectrum and so can be lower than the equivalent \LCDM value \citep{Rogers_S8tensionaxions}. The heavier axion ($m_{\rm{ax}}=10^{-21}\,\mathrm{eV}$) slightly boosts power on small scales relative to no axions (though the feedback still overall suppresses the power relative to the feedback-free limit). Feedback strength is therefore marginally under-estimated in order to account for the unmodelled axion effect. The lighter axion ($m_{\rm{ax}}=10^{-25}\,\mathrm{eV}$) strongly suppresses power. As this axion effect is also unmodelled, cosmological and feedback parameters are strongly biased from the truth. Neither case recovers the true cosmology as axions are not modelled.}
\label{fig:LCDM_cases}
\end{figure*}

In Fig.~\ref{fig:LCDM_cases}, we show two comparisons of marginalising with and without baryonic feedback for a mock LSST Y1 analysis of the shear correlation function (see Sec.~\ref{sec:data} for details on constructing the fake LSST data). In each comparison, we include both the effects of axions and feedback
in the data (using the models set out in Sec.~\ref{sec:modelling}), but ignore axions in the model used for inference. We therefore marginalise \(\Lambda\)CDM and systematics parameters (see Table \ref{tab:priors}), both with and without $\Theta_{\rm{AGN}}$.

For the heavier axion $m_{\rm{ax}}=10^{-21}\,\mathrm{eV}$, \(f_\mathrm{ax} = 0.1\) with extreme feedback \(\Theta_\mathrm{AGN} = 8.0\), the true value of $S_{8}$ is only recovered when marginalising $\Theta_{\rm{AGN}}$. Feedback suppresses the non-linear matter power spectrum, whilst the axion provides a slight boost of power on the same scales, but not more than the suppression (see \S~\ref{sec:feedback}). The result is that we recover $S_{8}$, but slightly underestimate feedback in order to account for the axion boost. When we do not account for feedback in the model, we instead find a bias to a low $S_{8}$, as has been seen in recent weak lensing analyses \citep{AAGPE2022,CPAAGPE}. The suppression of the matter power spectrum on small scales from feedback is interpreted as a low $S_{8}$.

For the lighter axion $m_{\rm{ax}}=10^{-25}\,\mathrm{eV}$, the power suppression caused by axions is so extreme (reducing power by up to 50\% on certain scales, as illustrated in Fig.~\ref{fig:Pk_suppression_plots}) that marginalising over baryonic feedback cannot capture the small-scale shape of the matter power spectrum. Cosmology is instead biased to low $S_{8}$ (even where the true $S_{8}$ is lower than the \LCDM value due to the effect of axions on the linear matter power spectrum). In this case, ignoring axions in the model results in overestimated feedback strength. $\Theta_{\rm{AGN}}$ attempts to compensate for the axion power suppression, hitting the upper prior limit at $\Theta_{\rm{AGN}}=9.3$. All together, without axions in the model, the axion effect masquerades either as spuriously high or low baryonic feedback.

\subsection{Extended cosmological analysis: marginalising over axions and feedback}\label{subsection:axionbaryon_fullextension}

\begin{figure*}
    \begin{minipage}{0.33\textwidth}
        \includegraphics[width=1.\columnwidth]{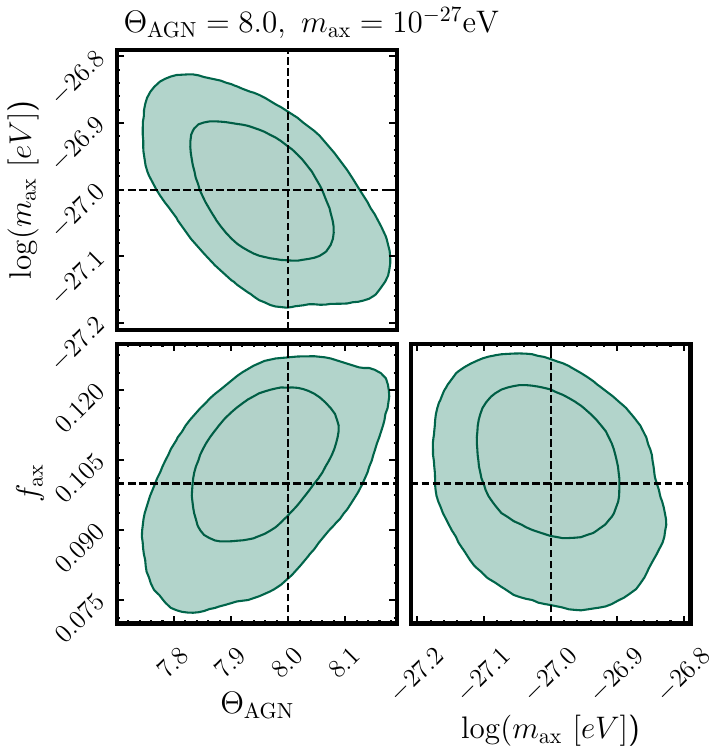} 
    \end{minipage}
    \begin{minipage}{0.33\textwidth}
        \includegraphics[width=1.\columnwidth]{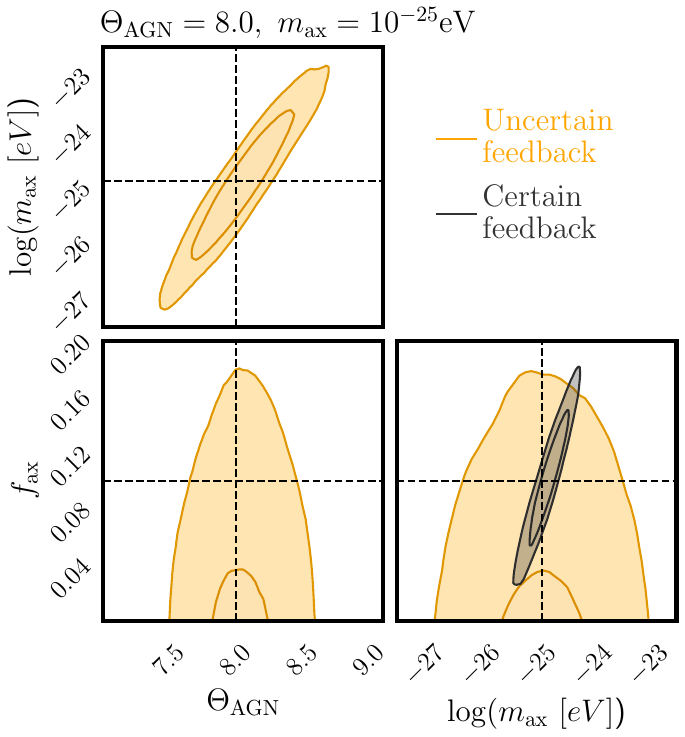} 
    \end{minipage}
    \begin{minipage}{0.33\textwidth}
        \includegraphics[width=1.\columnwidth]{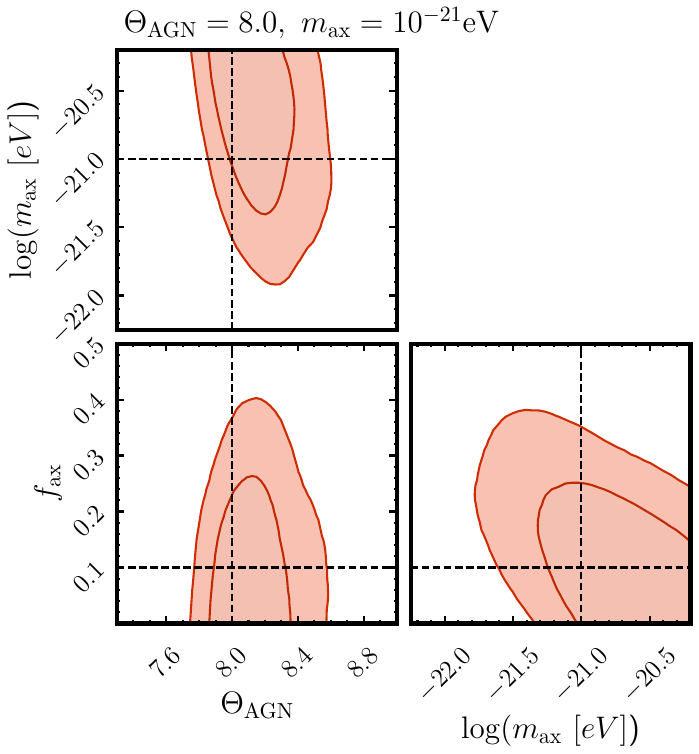} 
    \end{minipage}

\caption{
Posterior distributions from an extended cosmological analysis of a simulated LSST Y1-like shear correlation function, where axion mass, axion fraction and baryonic feedback parameters are varied. \textit{Left}: the fake data are generated for $m_{\rm{ax}}=10^{-27}\,\mathrm{eV}$, \(f_\mathrm{ax} = 0.1\) and \(\Theta_\mathrm{AGN} = 8.0\); \textit{centre}: $m_{\rm{ax}}=10^{-25}\,\mathrm{eV}$, \(f_\mathrm{ax} = 0.1\) and \(\Theta_\mathrm{AGN} = 8.0\); \textit{right}: $m_{\rm{ax}}=10^{-21}\,\mathrm{eV}$, \(f_\mathrm{ax} = 0.1\) and \(\Theta_\mathrm{AGN} = 8.0\). We report marginalised posteriors of \(\log (m_\mathrm{ax}\,[\mathrm{eV}])\), \(f_\mathrm{ax}\) and \(\Theta_\mathrm{AGN}\); the inner and outer contours respectively indicate the 68\% and 95\% credible regions. The true parameter values are shown as dashed black lines. For the centre panel, we also report the posterior assuming the feedback strength is known (black contours). For the lightest axion (\(m_\mathrm{ax} = 10^{-27}\,\mathrm{eV}\)), we recover the truth and a non-zero preference for axions. The heaviest axion (\(m_\mathrm{ax} = 10^{-21}\,\mathrm{eV}\)) has a weaker effect on the matter power spectrum for $k<10\,h\,\rm{Mpc}^{-1}$ and so we place only an upper limit $f_{\mathrm{ax}} \lesssim 0.4$, with no preference for axions; feedback strength is recovered. For the intermediate-mass axion (\(m_\mathrm{ax} = 10^{-25}\,\mathrm{eV}\)), the scales at which baryonic feedback and axions suppress the matter power spectrum are the same. It is therefore hard to disentangle the two effects without external information on feedback, although a competitive limit on the axion fraction is still feasible.}
\label{fig:axion_feedback_degeneracy}
\end{figure*}

In this extended cosmological analysis, we now vary all the parameters in Table~\ref{tab:priors}, i.e. \(\Lambda\)CDM, axion, feedback and systematics. In particular, we investigate the anticipated degeneracy between axion and baryonic feedback parameters. We show results in Fig.~\ref{fig:axion_feedback_degeneracy}.

For the lightest axion $m_{\rm{ax}}=10^{-27}\,\mathrm{eV}$, the scales at which axions and baryonic feedback suppress the power spectrum are distinct (see Fig.~\ref{fig:Pk_suppression_plots}). LSST Y1 shear can therefore place strong constraints on axion mass, axion fraction and feedback strength simultaneously, with a non-zero axion preference recovered. The posterior is consistent with the input truth. For this axion mass, axions and baryonic feedback can be distinguished in an extended cosmological analysis.

For the intermediate-mass axion $m_{\rm{ax}}=10^{-25}\,\mathrm{eV}$, the scales that axions and baryonic feedback suppress are roughly the same. This degeneracy limits the separation of axion and feedback effects in LSST Y1 shear. We therefore see a degeneracy between heavier axions (suppressing less) and stronger feedback (suppressing more). In the black contours in the centre panel of Fig.~\ref{fig:axion_feedback_degeneracy}, we show results where the strength of feedback is exactly understood, i.e., we fix \(\Theta_\mathrm{AGN}\) in the model. With such external information on feedback, we can recover a non-zero preference for axions, breaking the degeneracy between \(m_\mathrm{ax}\) and $\Theta_{\rm{AGN}}$. In practice, something intermediate to these cases is feasible, where we have an external prior on the feedback strength from feedback probes like X-rays and the Sunyaev-Zeldovich effect; we defer to future work a study on combining cosmic shear with other data. Nonetheless, even in the setting with no external information on feedback, a competitive limit on the axion fraction can be set with LSST cosmic shear alone (see Sec.~\ref{sec:conclusion} for more discussion). The full posteriors of cosmological, axion and feedback parameters in these intermediate-mass settings are shown in Appendix \ref{sec:app_B}.

For the heaviest axion $m_{\rm{ax}}=10^{-21}\,\mathrm{eV}$, a small boost in power is seen on scales $k \sim 1\,h\,\mathrm{Mpc}^{-1}$ (see Fig.~\ref{fig:Pk_suppression_plots} and discussion in Sec.~\ref{sec:feedback}). However, $\Theta_{\mathrm{AGN}}$ is mildly degenerate with $m_{\mathrm{ax}}$, with a boost in power from axions in opposition to suppression from baryonic feedback. As a result, we only set an upper limit on the axion fraction. While this is not a detection of the axion, such a limit is still useful, since large axion fractions at this mass have been suggested to explain the ``small-scale crisis'' in cosmology \citep{Hu:2000,Marsh_cuspcore, Bullock_smallscale}. The effects of baryonic feedback will likely erase FDM features in the matter power spectrum for $k>10\,h\,\rm{Mpc}^{-1}$. LSST Y1 does not have sensitivity to this regime in any case (see Sec.~\ref{sec:reconstruction}).

\subsection{Matter power spectrum reconstruction}
\label{sec:reconstruction}

Weak lensing data can be used to constrain the shape of the matter power spectrum \citep{Tegmark_Zaldarriaga, Chabanier:2019b, DouxDESHarmonic}. \citet{CPAAGPE} perform a parametric reconstruction of the non-linear matter power spectrum, finding evidence using Dark Energy Survey (DES) Year 3 and Kilo Degree Survey (KiDS) data \citep{Asgari:2021,Secco:2022,amon:2022} for suppression at $k>0.1\,h\,\rm{Mpc}^{-1}$ compared to the \Planck best-fit \LCDM cosmology. Recent works have also tried different parameterisations to fit the shape of the matter power spectrum with weak lensing data \citep{Broxterman_Kuijken, Silvestri, Sarmiento}, finding consistent results. \citet{Preston_2024} demonstrate the ability of future weak lensing data to reconstruct the non-linear matter power spectrum. We outline the reconstruction method below (Sec.~\ref{sec:reconstruction_method}), using it to forecast the LSST Y1 shear reconstruction of the non-linear matter power spectrum in cosmologies with axions and baryonic feedback (Sec.~\ref{sec:reconstruction_forecast}).

\subsubsection{Reconstruction method}
\label{sec:reconstruction_method}

First, we fix the \LCDM parameters ($\Omega_{\rm{m}}$, $\Omega_{\rm{b}}$, $A_{s}$, $n_{s}$, $h$) to their \Planck best-fit \LCDM values (Table \ref{tab:priors}). This fixes the primordial power spectrum and the subsequent background expansion history. We then vary the shape of the non-linear matter power spectrum at \(z = 0\). We achieve this by varying the amplitude of the \(z = 0\) non-linear matter power spectrum \(P_\mathrm{m}(k, z=0)\) relative to the \texttt{EuclidEmulator} \(z = 0\) non-linear matter power spectrum given the \textit{Planck} best-fit \(\Lambda\)CDM cosmology \(P_\texttt{EucEmu}^{\Lambda\mathrm{CDM}}(k, z=0)\). We vary this ratio in eight wavenumber bins of equal width in the interval $-1 \leq \log (k\,[h\,\mathrm{Mpc}^{-1}]) \leq 1$, plus two extra bins that each capture all modes greater and smaller than this interval, with bin centres \(k_i\):
\begin{equation}
   \hat{\mathcal{P}}(k_{i}) = \frac{P_{\rm{m}}(k_i,z=0)}{P_\texttt{EucEmu}^{\Lambda\mathrm{CDM}}(k_i, z=0)}.
\label{equ:BinningPkz}
\end{equation}
In order to compare to cosmic shear, which is sensitive to the \(z > 0\) matter power spectrum (Eq.~\eqref{equ:Ckappa1}), we then assume \LCDM growth of structure from $z=2$ to the present day, where weak lensing is most sensitive to the power spectrum at $z\lesssim0.5$. This is a good approximation for LSST sensitivity and is discussed more in Sec.~3 of \citet{Preston_2024}. We infer \(\hat{\mathcal{P}}(k_i)\) given forecast LSST Y1 cosmic shear data, marginalising over the systematics parameters in Table \ref{tab:priors}, i.e., uncertainties in intrinsic alignments, the source redshift distribution and shear calibration. We do not vary axion or feedback parameters as we are not here inferring a specific cosmological model but rather are inferring the matter power spectrum. This procedure exploits the small scales probed by weak lensing, recovering $P_{\rm{m}}(k)$ up to $k \sim 10\,h\,\mathrm{Mpc}^{-1}$, without relying on cosmic shear as a probe of the background expansion history. 

\begin{figure*}
    \begin{minipage}{0.33\textwidth}
    \includegraphics[width=1.\columnwidth]{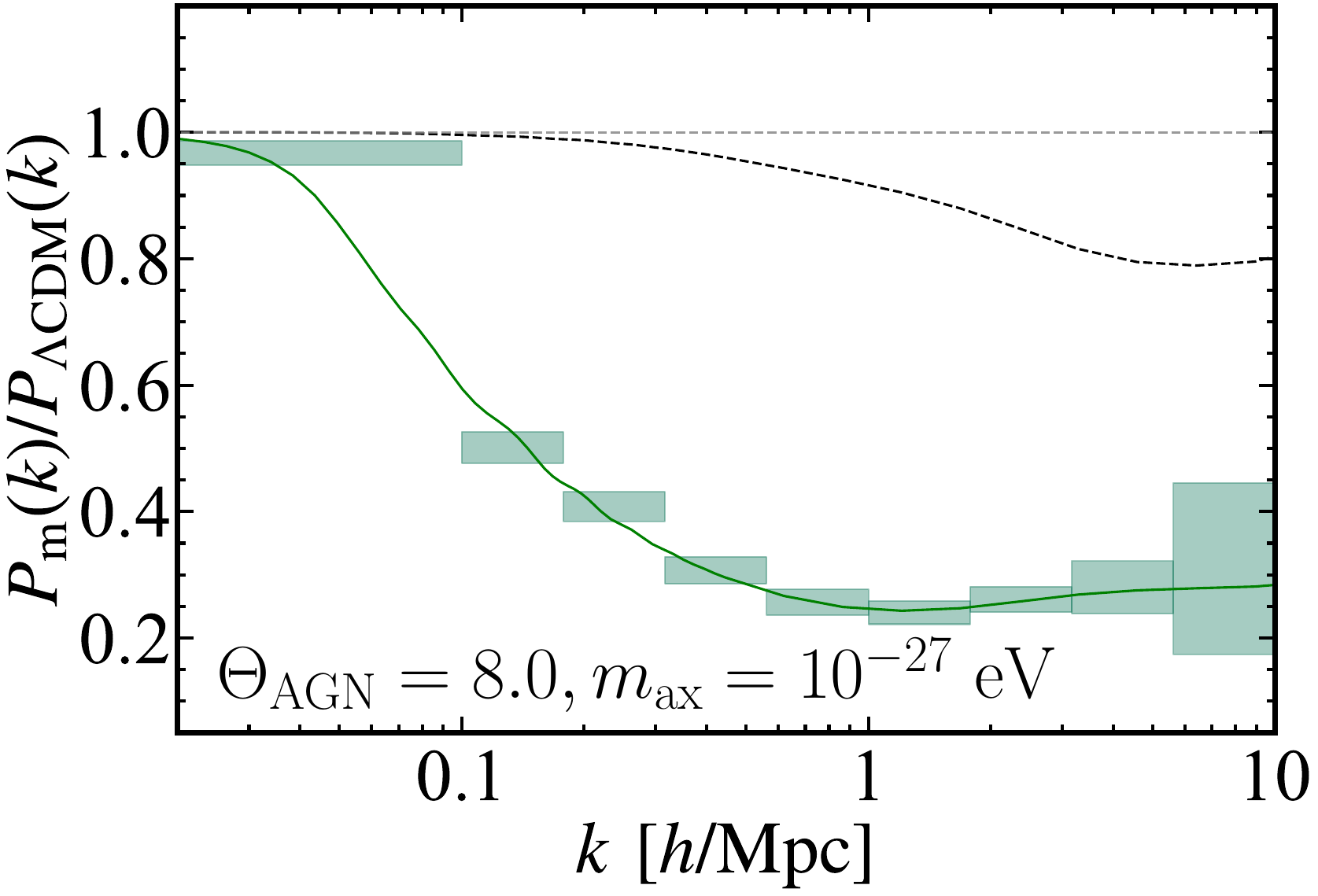} 
    \end{minipage}
    \begin{minipage}{0.33\textwidth}
        \includegraphics[width=1.\columnwidth]{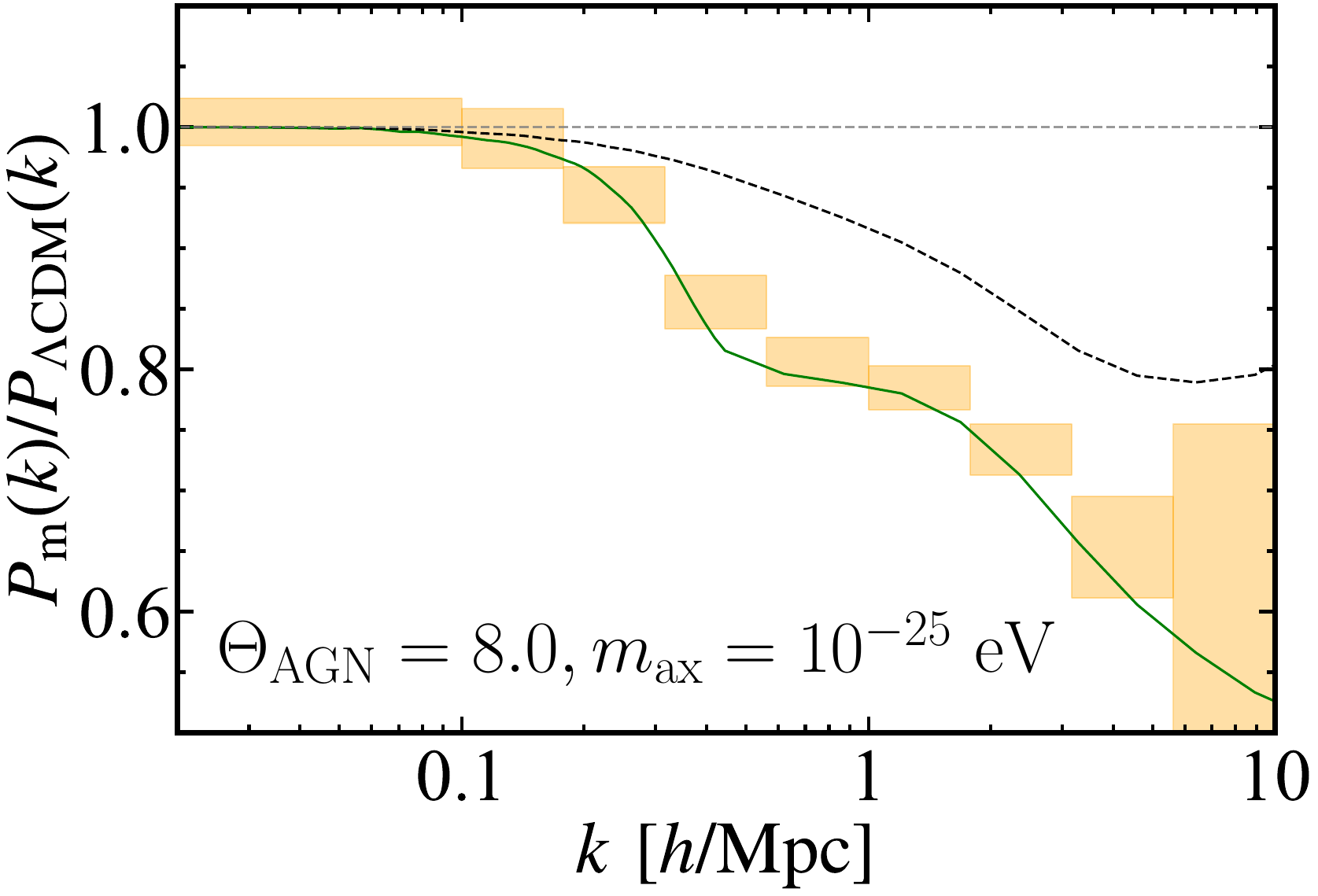} 
    \end{minipage}
    \begin{minipage}{0.33\textwidth}
        \includegraphics[width=1.1\columnwidth]{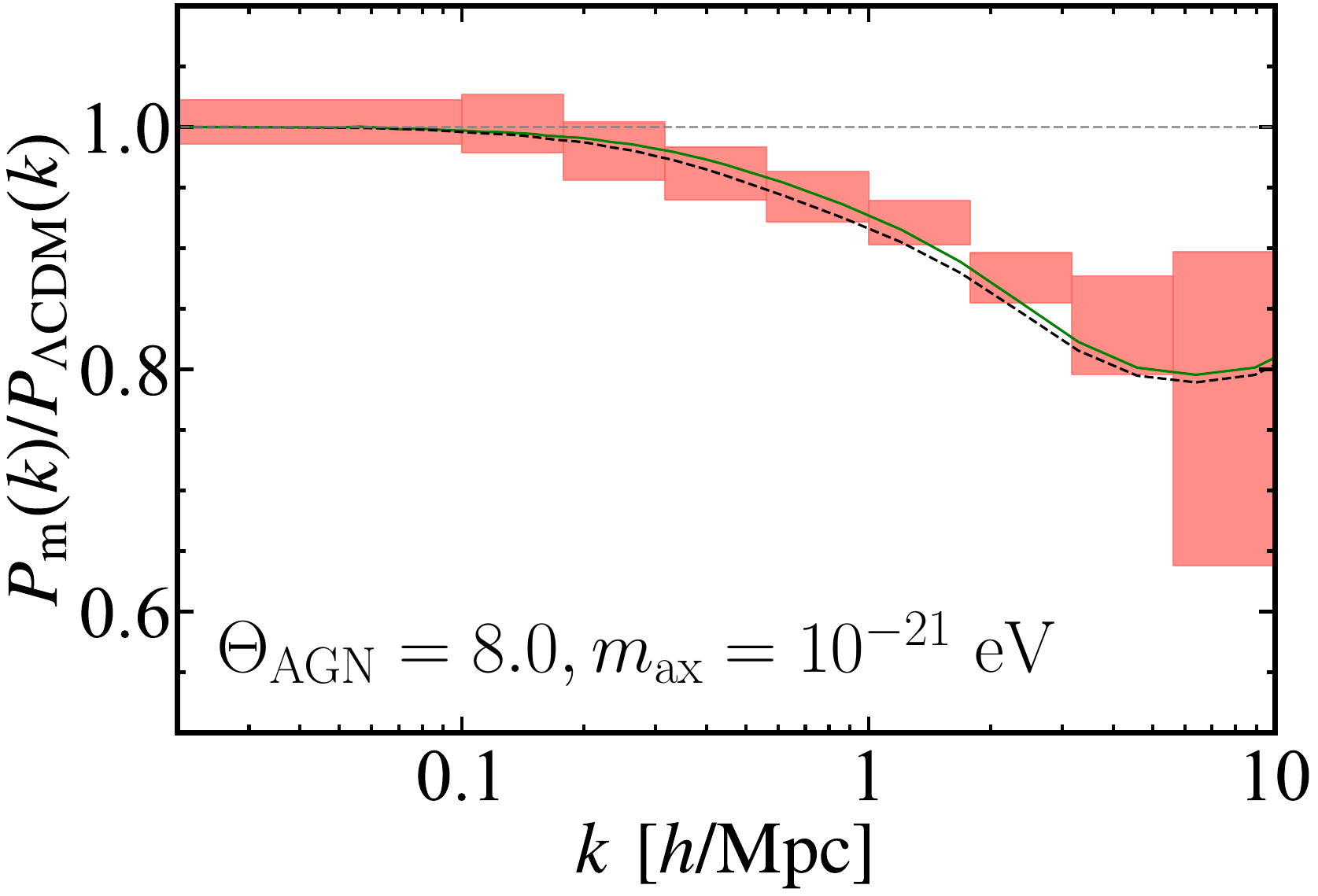} 
    \end{minipage}

\caption{Forecast LSST Y1 shear inference of the reconstructed \(z=0\) non-linear matter power spectrum \(P_\mathrm{m}(k)\), normalised by the feedback-free \textit{Planck} \(\Lambda\)CDM cosmology from \texttt{EuclidEmulator}. \textit{From left to right}, the fake data are the same as in Fig.~\ref{fig:axion_feedback_degeneracy}. The width of each box indicates the width of the wavenumber bin \(i\); the height indicates the 68\% credible interval centred at the mean posterior. The true power spectrum is the solid green line. In each case, we recover the true power suppression shape. We do not plot the largest \(k\) bin, as this is largely prior dominated and we have little sensitivity with cosmic shear to scales $k > 10\,h\,\mathrm{Mpc}^{-1}$ at LSST Y1 sensitivity. The feedback-only suppression for $\Theta_{\mathrm{AGN}}=8.0$ is plotted in each panel in dotted black and is sometimes distinguished from the combined axion and feedback scenario, i.e., we can distinguish axions from CDM at \(m_\mathrm{ax} \sim 10^{-25}\,\mathrm{eV}\) if we have external feedback information.}
\label{fig:Pk_reconstruction_plots}
\end{figure*}

Having inferred the matter power spectrum, it is then a subsequent task to identify the astrophysical and/or cosmological processes responsible for the shape of the matter power spectrum. The power of this approach is that the inferred shape of the non-linear matter power spectrum is agnostic to specific feedback or dark matter models (or other previously unaccounted effects on the power spectrum) during the inference. Ultimately, reconstructing the matter power spectrum is then a diagnostic tool to identify any deviations from the \Planck \LCDM cosmology. Rather than testing many different extensions to \(\Lambda\)CDM, power spectrum reconstruction identifies features on small scales ($0.1\lesssim k~[h\,\rm{Mpc}^{-1}]\lesssim10$) that viable extensions should satisfy.

\subsubsection{Forecast reconstruction from LSST Y1 cosmic shear}
\label{sec:reconstruction_forecast}

In Fig.~\ref{fig:Pk_reconstruction_plots}, we reconstruct the non-linear matter power spectrum given forecast LSST Y1 cosmic shear for three different axion scenarios with baryonic feedback. These three scenarios are the same as we consider in Sec.~\ref{subsection:axionbaryon_fullextension}. We recover the shapes of all three target power spectra within the 68\% credible intervals.\footnote{The lowest $k$ bin contains all modes $k<0.1\,h\,\mathrm{Mpc}^{-1}$. Most of the constraining power in this bin comes from the highest $k$ modes within the bin.} This is true even for the combined axion and baryonic feedback scenario in the centre panel, which features a slight plateau at $k \sim 1\,h\,\rm{Mpc}^{-1}$. This more complex behaviour results from the axion soliton (see  Fig.~\ref{fig:Pk_suppression_plots} and discussion in Sec.~\ref{sec:feedback}). For the lightest axion $m_{\rm{ax}}=10^{-27}\,\mathrm{eV}$, the power spectrum reconstruction captures the input extreme suppression. No baryonic feedback model currently predicts such an extreme suppression. Detection of such an extreme suppression by reconstruction would provide evidence of either extreme feedback beyond predictions from current hydrodynamical simulations or beyond-\LCDM physics. We however note that such a strong axion suppression is already inconsistent with \textit{Planck} CMB and Baryon Oscillation Spectroscopic Survey (BOSS) galaxy clustering data \citep{Rogers_S8tensionaxions}.

We also show the feedback-only scenario. For the intermediate-mass axion \((m_\mathrm{ax} = 10^{-25}\,\mathrm{eV})\), the combined axion and feedback suppression is distinguishable from the feedback-only limit. This is consistent with the result in Sec.~\ref{subsection:axionbaryon_fullextension} that axions of this mass can be distinguished from pure CDM, if we have external feedback information. In practice, it is feasible to obtain a prior on feedback strength from complementary observations, i.e., to restrict the range of viable feedback suppressions and then to search for additional suppression from dark matter physics. The power of reconstruction is to identify the wavenumbers at which deviation from CDM is viable or not. For all three scenarios, we forecast that LSST Y1 shear can distinguish from the feedback-free \(\Lambda\)CDM limit, consistent with the results in Sec.~\ref{subsection:axionbaryon_fullextension}.

\begin{figure}
\includegraphics[width=1.\columnwidth]{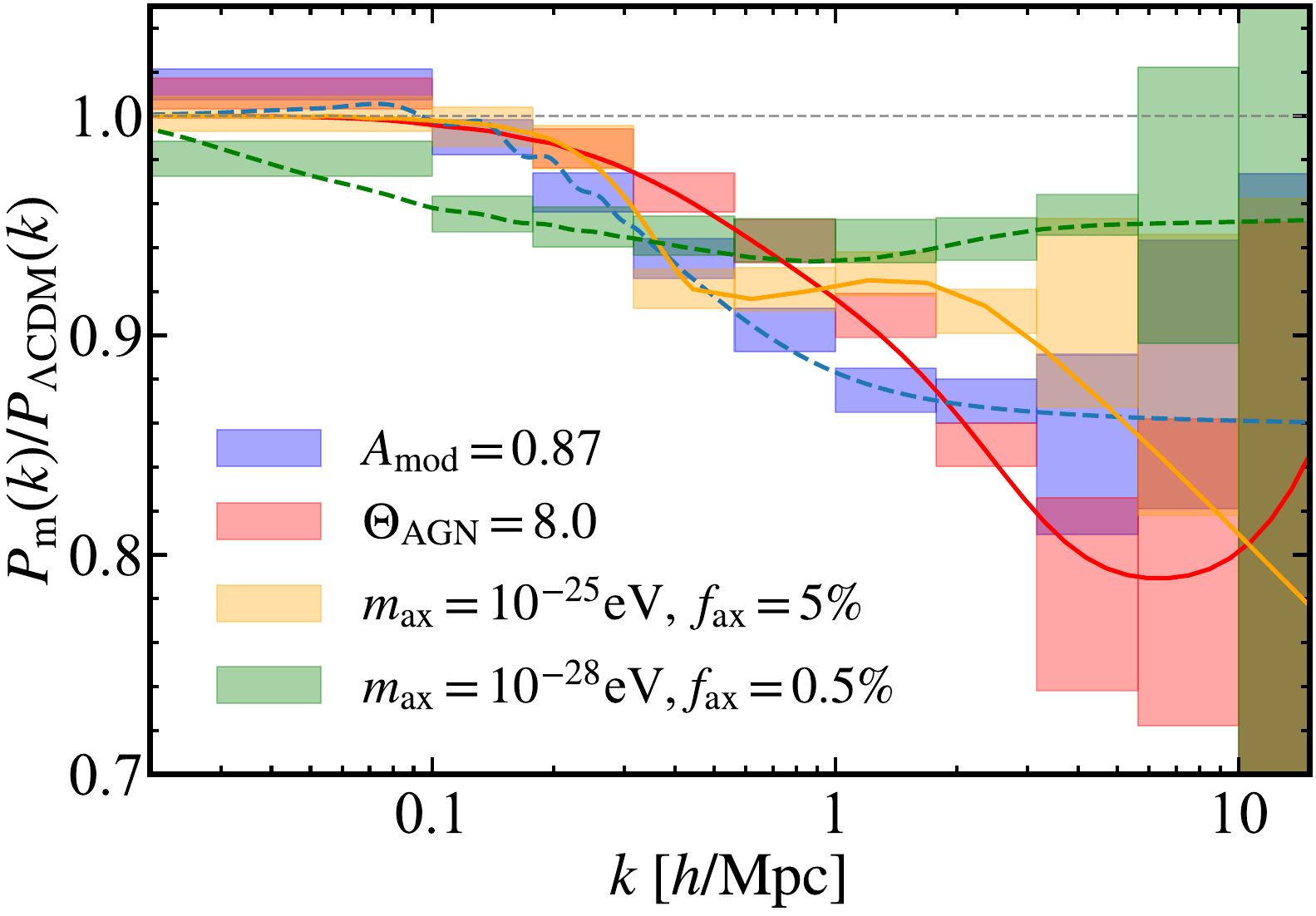} 
\caption{As Fig.~\ref{fig:Pk_reconstruction_plots}, but for the power suppression scenarios introduced in Fig.~\ref{fig:toy_example}. The solid and dashed lines are the true power spectra. LSST Y1 shear reconstruction of the matter power spectrum is forecast to distinguish between these scenarios, whereas a \LCDM inference of $S_8$ and $\Omega_{\rm{m}}$ cannot (Fig.~\ref{fig:toy_example}).}
\label{fig:S8_compilation_Pkreconstruction}
\end{figure}

In Fig.~\ref{fig:S8_compilation_Pkreconstruction}, we reconstruct the non-linear matter power spectrum for the four power suppression scenarios we introduced in Fig.~\ref{fig:toy_example}. Each scenario leads to nearly identical constraints in the $S_8 - \Omega_\mathrm{m}$ plane given LSST Y1 shear and assuming \(\Lambda\)CDM. The LSST Y1 power spectrum reconstructions clearly differentiate the scenarios without any requirement to model explicitly the physics causing the suppression. This ability to distinguish between \(\Lambda\)CDM + feedback models and feedback-free axion models is consistent with the results in Sec.~\ref{subsection:axionbaryon_fullextension}. Fig.~\ref{fig:S8_compilation_Pkreconstruction} does not address how to differentiate between axions and \(\Lambda\)CDM both in the presence of feedback (see Sec.~\ref{subsection:axionbaryon_fullextension} for these results). This example instead highlights the power of power spectrum reconstruction in providing a complementary way of analysing weak lensing data beyond inferring only cosmological parameters like \(S_8\) and \(\Omega_\mathrm{m}\). It is in effect a diagnostic tool to identify which scales, if any, are discrepant from the \(\Lambda\)CDM expectation, while \(S_8\) inferences throw away this information.

\section{Discussion and conclusions}\label{sec:conclusion}

This paper has investigated the prospects of using weak lensing data to detect a non-standard dark matter component. We highlighted the degeneracies between axions, baryonic feedback and cosmology, emphasizing the need for accurate modelling of small cosmological scales (\(k > 1\,h\,\mathrm{Mpc}^{-1}\)) in order to extract unbiased results from weak lensing measurements.

In Sec.~\ref{sec:modelling}, we introduce a heuristic mixed dark matter halo model for computing power spectra in which a fraction \(f_\mathrm{ax}\) of the dark matter consists of an axion-like particle of mass $m_{\rm{ax}}$ [Eq.~\eqref{equ:total_pk}]. This model is approximate since baryonic feedback is assumed to affect axions and CDM in the same way via the third term in Eq.~\eqref{equ:total_pk}, i.e., that \(\Theta_\mathrm{AGN}\) is uncorrelated with \(m_\mathrm{ax}\) and \(f_\mathrm{ax}\). We argue that this is a good approximation as the feedback model is calibrated on cosmologies with massive neutrinos, which are qualitatively similar to the axion models that we consider in their small-scale power suppression. In any case, the physics of baryonic feedback is uncertain already under \(\Lambda\)CDM. The model that we construct should therefore be regarded as a first step towards understanding the degeneracies between axions and baryonic feedback. Ultimately, detailed hydrodynamical simulations of baryon-CDM-axion cosmologies will be required to model the matter power spectrum (either to construct an emulator or to calibrate a halo model) with sufficient accuracy for data analysis.

Here, we investigate extensions to \LCDM including both baryonic feedback and axions. It is already clear from previous work \citep{AAGPE2022} that in a baryonic feedback + \LCDM cosmology, failure to sufficiently account for baryonic feedback will bias cosmological inference. A feedback model must be flexible enough to model strong feedback scenarios, as suggested by both kinetic and thermal Sunyaev-Zeldovich  measurements \citep[see e.g.][] {Bigwood:2024,Efstathiou:2025}. This remains true in the extension with axions that we consider here, hence the extended prior on \(\Theta_\mathrm{AGN}\) that we use (Table \ref{tab:priors}).

In Sec.~\ref{sec:ignore_axions}, we study the impact of ignoring axions in the cosmological model if they do exist in the data. For the heavier axion that we consider $m_{\rm{ax}} = 10^{-21}\,\mathrm{eV}$, our LSST weak lensing forecast shows that marginalising over baryonic feedback strength while treating all of the dark matter as CDM will yield unbiased constraints in the  $S_8-\Omega_\mathrm{m}$ plane at the expense of underestimating feedback strength. For axions lighter than $10^{-21}\,\mathrm{eV}$, it is critical to marginalise over axion and feedback parameters to avoid biases to cosmological parameters. In this setting, feedback strength will be overestimated but still unable to account for the additional suppression from axions.

In Sec.~\ref{subsection:axionbaryon_fullextension}, we extend the analysis to model weak lensing two-point statistics with varying axion mass, axion fraction and feedback strength, in addition to \LCDM cosmological parameters and systematic uncertainties in the data and model (intrinsic alignments, source redshifts, shear calibration). This analysis shows that, for $10^{-27}<m_{\rm{ax}}\,[\mathrm{eV}]\leq10^{-21}$, the picture is complex. Scales of power suppression due to axions and baryonic feedback are similar and become difficult to disentangle without external information on the strength of feedback. Constraints on feedback from e.g. the kinetic and thermal Sunyaev-Zeldovich (SZ) effects \citep{Bigwood:2024,Efstathiou:2025} and X-ray observations of gas in galaxy groups and clusters \citep{schneider:2021,Ferreira:2023syi} will be necessary to detect axions in this mass range. This direction is already being investigated. With such knowledge of baryonic physics, the middle panel of Fig.~\ref{fig:axion_feedback_degeneracy} suggests we could find a preference for an axion of \(m_\mathrm{ax} = 10^{-25}\,\mathrm{eV}\) being 10\% of the dark matter at roughly $3\sigma$ significance. Nonetheless, even without external information, LSST cosmic shear alone will set competitive limits on the axion fraction (see below). The degeneracy between baryonic feedback and axion mass in this mass range determines the regime where full hydrodynamical axion simulations can be targeted.

For the lightest axion that we consider $m_{\rm{ax}}=10^{-27}\,\mathrm{eV}$, axion physics leads to power suppression on scales that are larger than those affected in any plausible model of baryonic feedback. In this mass regime, the effects of axions and baryonic feedback can be distinguished in cosmic shear \citep[although see existing results from the CMB and galaxy clustering,][]{Rogers_S8tensionaxions}. Although we do not explicitly test for axions heavier than \(10^{-21}\,\mathrm{eV}\), we conclude that LSST Y1 shear will have sensitivity to heavier axions if the modelling challenge can be met, since we can still rule out large axion fractions at \(10^{-21}\,\mathrm{eV}\). We defer to future work a detailed study of this regime coupled with more sophisticated simulations. We also do not explicitly test for axions lighter than \(10^{-27}\,\mathrm{eV}\) as we then must account for the transition from dark energy-like to dark matter-like in the formation of axion halos which is beyond our current modelling capability.

In Sec.~\ref{sec:reconstruction}, we investigate matter power spectrum reconstruction from weak lensing in the presence of axions and feedback following the approach of \citet{Preston_2024}. With high signal-to-noise weak lensing data, which LSST Y1 will provide, we find that the matter power spectrum can be reconstructed with high accuracy and precision (limited by systematics such as intrinsic alignments). As shown in Fig.~\ref{fig:Pk_reconstruction_plots}, power spectrum reconstruction serves as a powerful diagnostic tool in distinguishing between baryonic feedback and axions.

Some recent weak lensing results have suggested that the linear clumpiness parameter \(S_8\) is systematically lower than expected from CMB data \citep[``\(S_8\) tension'',][]{Amon:2021,dalal2023hyper,li2023hyper}. \citet{AAGPE2022} suggest that unmodelled extreme baryonic feedback could be biasing \(S_8\) low; \citet{Rogers_S8tensionaxions} point out that \(S_8\) really could be lower owing to axion Jeans suppression. In Fig.~\ref{fig:S8_compilation_Pkreconstruction}, we find that LSST Y1 cosmic shear matter power spectrum reconstruction could differentiate between pure baryonic feedback and axion suppression. In practice, we expect always some feedback effect and so we advocate for dedicated hydrodynamical axion simulations in order to understand their interplay in detail. Power spectrum reconstruction is thus complementary to the approach of inferring dark matter and feedback models directly from cosmic shear two-point statistics. First, it identifies at which scales dark matter and feedback effects could be differentiated. Second, since the true dark matter model (CDM, axions, warm dark matter, interacting dark matter, etc.) is not \textit{a priori} known, then reconstruction helps identify what features viable models should satisfy, i.e., it is a diagnostic tool.

\begin{figure}
    \begin{minipage}{\columnwidth}
    \includegraphics[width=1.\columnwidth]{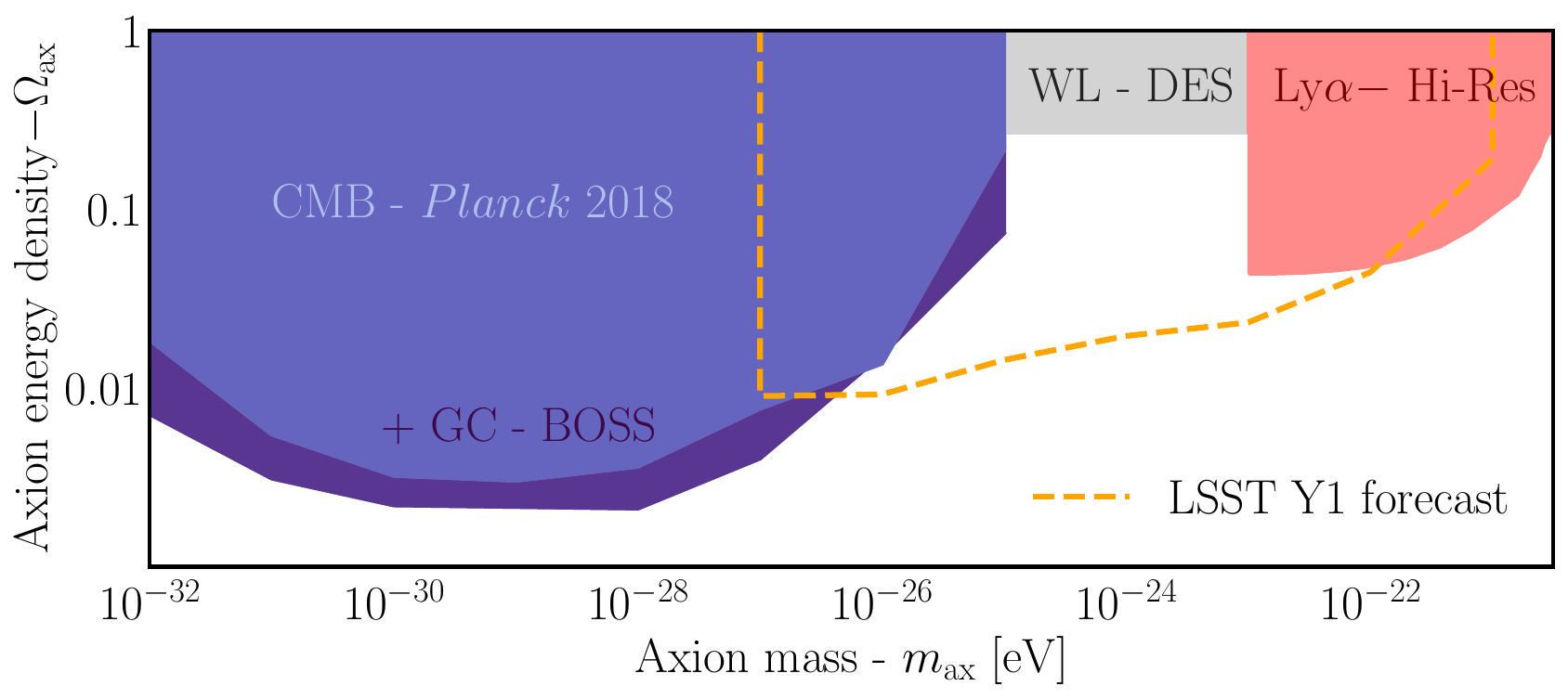} 
    \end{minipage}
\caption{Compilation of current exclusions (\textit{solid regions}) in the ultra-light axion parameter space, alongside the forecast sensitivity of LSST Y1 cosmic shear (region within the \textit{gold dashed line}). Lighter blue and darker blue regions respectively show the parameter space ruled out by \textit{Planck} 2018 CMB alone and in combination with Baryon Oscillation Spectroscopic Survey (BOSS) galaxy clustering (GC) \citep{Rogers_S8tensionaxions}. The red region shows the parameter space ruled out by high-resolution observations of the Lyman-$\alpha$ forest (Ly\(\alpha\)) \citep{Kobayashi_Lyman_DM,Rogers_Peiris_Lyman}. The grey region shows the parameter space currently ruled out by Dark Energy Survey (DES) weak lensing (WL) \citep{Dentler_Axions_weaklensing}. All current and forecast exclusions are shown at 95\% c.l. See text for more details on how we construct the LSST forecast. We restrict ourselves here to limits using axion Jeans suppression in different tracers of the matter power spectrum; see also e.g. \citet{Winch_Rogers_etal} for limits in this mass range.}
\label{fig:axion_compilation}
\end{figure}

\begin{table}
    \caption{Forecast LSST Y1 shear 95\% c.l. upper limits on axion fraction $f_{\mathrm{ax}}$ (\textit{right}) at different axion masses $m_{\mathrm{ax}}$ (\textit{left}; see also Fig.~\ref{fig:axion_compilation}). See text for more details.}
\label{tab:LSSTY1_forecast}
\begin{center}
\begin{tabular}{ll}
$\log(m_{\rm{ax}}\,[\mathrm{eV]})$ & $f_{\mathrm{ax}}$ (95\% c.l.) \tabularnewline
\hline 

 ~~~~~ -27 &~~~ $< 3.0\%$ \tabularnewline
 ~~~~~ -26 &~~~ $< 3.1\%$\tabularnewline
 ~~~~~ -25 &~~~ $< 4.7\%$
\tabularnewline
 ~~~~~ -24 &~~~ $< 6.5\%$ \tabularnewline
 ~~~~~ -23 &~~~ $< 7.7\%$
 \tabularnewline
 ~~~~~ -22 &~~ $< 14.8\%$ \tabularnewline
 ~~~~~ -21 &~~ $< 63.7\%$ \tabularnewline

\hline 
\end{tabular}
\end{center}
\end{table}

In Fig.~\ref{fig:axion_compilation}, we show how cosmic shear provides a competitive and powerful probe of underexplored regions of axion mass and energy density parameter space. We summarise the LSST Y1 cosmic shear forecast in Table \ref{tab:LSSTY1_forecast}. To make the LSST Y1 forecast, we construct CDM-like fake data and then infer the 95\% upper limit on \(f_\mathrm{ax}\), for different fixed values of \(m_\mathrm{ax}\) while also fixing \(\Theta_\mathrm{AGN}\), thus forming the gold dashed line in Fig.~\ref{fig:axion_compilation}. We otherwise marginalise all \(\Lambda\)CDM and systematics parameters (see Table \ref{tab:priors}). We restrict this forecast to $10^{-27} \leq m_{\rm{ax}}\,[\mathrm{eV}]\leq10^{-21}$ since we do not currently have the modelling capability outside this range (see above). By fixing \(\Theta_\mathrm{AGN}\), we are approximating the limit where we have external information on feedback strength (SZ, X-rays). As a consequence, this forecast should be approximately taken as a best-case limit if the axion and feedback modelling is improved.

While CMB and galaxy clustering data rule out large ULA fractions for \(m_\mathrm{ax} \lesssim 10^{-26}\,\mathrm{eV}\) \citep{Hlozek_CMB_axions2015,Hlozek_CMB_axions2018,Lague_axions_galaxyclustering,Rogers_S8tensionaxions} and the Lyman-\(\alpha\) forest \citep{Irsic_Lyman_DM,Kobayashi_Lyman_DM,Rogers_Peiris_Lyman} sets limits for \(m_\mathrm{ax} \gtrsim 10^{-23}\,\mathrm{eV}\) \citep[see also][]{Rogers_Poulin}, there is an under-explored intermediate mass regime. A previous analysis of 3 \(\times\) 2-pt cosmic shear and galaxy clustering from DES \citep{Dentler_Axions_weaklensing} rules out axions contributing $100\%$ of the dark matter in this intermediate regime. \citet{Dentler_Axions_weaklensing} do not consider the mixed cold and axion dark matter scenario. We forecast that a future cosmic shear survey like LSST Y1 can be competitive with the CMB at constraining $m_{\mathrm{ax}}=10^{-27}$eV axions, with sensitivity possible down to a 3\% contribution (95\% c.l.). For $m_{\mathrm{ax}}=10^{-25}\,\mathrm{eV}$, an LSST Y1-like experiment can improve on previous constraints from DES, placing a limit $f_{\mathrm{ax}}<5\%$ (95\% c.l.). \citet{Winch_Rogers_etal} also place constraints on axion mass and fraction in this regime using the galaxy ultraviolet luminosity function measured by the Hubble and James Webb Space Telescopes and information from \textit{Planck}. This method places constraints competitive with cosmic shear and is dependent on modelling the HMF and galaxy-halo connection for $z>4$.

An observable to investigate in the future is the auto- and cross-correlations of cosmic shear and galaxy clustering in combination as a $3\times2\mathrm{-pt}$ statistic. The additional information supplied by galaxy clustering and galaxy-galaxy lensing would likely improve parameter constraints beyond what we report here for cosmic shear alone. However, in using galaxy clustering, we require a robust galaxy bias model that incorporates axions to small scales. \citet{Rogers_S8tensionaxions} model galaxy bias in the presence of axions for \(k \leq 0.4\,h\,\mathrm{Mpc}^{-1}\).

In conclusion, we find that, despite an anticipated degeneracy between axion Jeans suppression and baryonic feedback, cosmic shear by itself can set competitive limits on the axion mass and energy density in an under-explored part of parameter space. With external information on the strength of feedback, from e.g. the Sunyaev-Zeldovich effect or X-ray observations \citep{schneider:2021,Ferreira:2023syi,Bigwood:2024}, sensitivity to axions will increase further and it is feasible for a stage-IV survey like LSST Y1 to detect a 10\% contribution of axions with \(m_\mathrm{ax} = 10^{-25}\,\mathrm{eV}\) at \(\sim 3 \sigma\) significance. We defer to future work a detailed study combining shear with other probes like the CMB. While we have considered the specific setting of LSST Y1, our general conclusions apply to other weak lensing surveys like \textit{Euclid} \citep{EuclidForecast} and the Roman Space Telescope \citep{Roman}, as well as CMB lensing probes like the Atacama Cosmology Telescope \citep{ACT_primary_CMB}, South Pole Telescope \citep{SPT_cosmology}, Simons Observatory \citep{SimonsForecast,SimonsObservatory:2025wwn}, CMB-S4 \citep{CMB_S4_snowmass,Dvorkin:2022bsc} and the proposed CMB-HD experiment \citep{CMB_HD_snowmass}. In parallel with preparations for standard cosmological analyses with these observatories, we advocate for detailed hydrodynamical axion simulations in order to understand the interplay between feedback and novel dark matter physics.

\section*{Acknowledgements}
CP is supported by a UKRI Science and Technology Facilities Council studentship. KKR is supported by an Ernest Rutherford Fellowship from the UKRI Science and Technology Facilities Council (grant number ST/Z510191/1). GPE thanks the Leverhulme Foundation for the award of an Emeritus Fellowship.

\section*{Data availability}
The forecast in this work is generated following the \textit{Rubin} LSST Dark Energy Science Collaboration Science Requirements Document \citep{DESC_requirements}.

\bibliographystyle{mnras} 
\bibliography{axionTagn}

\begin{thebibliography}{}
\makeatletter
\relax
\def\mn@urlcharsother{\let\do\@makeother \do\$\do\&\do\#\do\^\do\_\do\%\do\~}
\def\mn@doi{\begingroup\mn@urlcharsother \@ifnextchar [ {\mn@doi@} {\mn@doi@[]}}
\def\mn@doi@[#1]#2{\def\@tempa{#1}\ifx\@tempa\@empty \href {http://dx.doi.org/#2} {doi:#2}\else \href {http://dx.doi.org/#2} {#1}\fi \endgroup}
\def\mn@eprint#1#2{\mn@eprint@#1:#2::\@nil}
\def\mn@eprint@arXiv#1{\href {http://arxiv.org/abs/#1} {{\tt arXiv:#1}}}
\def\mn@eprint@dblp#1{\href {http://dblp.uni-trier.de/rec/bibtex/#1.xml} {dblp:#1}}
\def\mn@eprint@#1:#2:#3:#4\@nil{\def\@tempa {#1}\def\@tempb {#2}\def\@tempc {#3}\ifx \@tempc \@empty \let \@tempc \@tempb \let \@tempb \@tempa \fi \ifx \@tempb \@empty \def\@tempb {arXiv}\fi \@ifundefined {mn@eprint@\@tempb}{\@tempb:\@tempc}{\expandafter \expandafter \csname mn@eprint@\@tempb\endcsname \expandafter{\@tempc}}}

\bibitem[\protect\citeauthoryear{{Abazajian} et~al.,}{{Abazajian} et~al.}{2022}]{CMB_S4_snowmass}
{Abazajian} K.,  et~al., 2022, \mn@doi [arXiv e-prints] {10.48550/arXiv.2203.08024}, \href {https://ui.adsabs.harvard.edu/abs/2022arXiv220308024A} {p. arXiv:2203.08024}

\bibitem[\protect\citeauthoryear{Abbott \& Sikivie}{Abbott \& Sikivie}{1983}]{Abbott_invisibleaxion}
Abbott L.~F.,  Sikivie P.,  1983, \mn@doi [Phys. Lett. B] {10.1016/0370-2693(83)90638-X}, 120, 133

\bibitem[\protect\citeauthoryear{{Adame} et~al.,}{{Adame} et~al.}{2025}]{DESI_BAO_2024}
{Adame} A.~G.,  et~al., 2025, \mn@doi [\jcap] {10.1088/1475-7516/2025/02/021}, \href {https://ui.adsabs.harvard.edu/abs/2025JCAP...02..021A} {2025, 021}

\bibitem[\protect\citeauthoryear{{Ade} et~al.,}{{Ade} et~al.}{2019}]{SimonsForecast}
{Ade} P.,  et~al., 2019, \mn@doi [\jcap] {10.1088/1475-7516/2019/02/056}, \href {https://ui.adsabs.harvard.edu/abs/2019JCAP...02..056A} {2019, 056}

\bibitem[\protect\citeauthoryear{{Akerib} et~al.,}{{Akerib} et~al.}{2022}]{WIMP_SNOWMASS}
{Akerib} D.~S.,  et~al., 2022, \mn@doi [arXiv e-prints] {10.48550/arXiv.2203.08084}, \href {https://ui.adsabs.harvard.edu/abs/2022arXiv220308084A} {p. arXiv:2203.08084}

\bibitem[\protect\citeauthoryear{{Alam} et~al.,}{{Alam} et~al.}{2021}]{Alam:2021a}
{Alam} S.,  et~al., 2021, \mn@doi [\prd] {10.1103/PhysRevD.103.083533}, \href {https://ui.adsabs.harvard.edu/abs/2021PhRvD.103h3533A} {103, 083533}

\bibitem[\protect\citeauthoryear{Amendola \& Barbieri}{Amendola \& Barbieri}{2006}]{Amendola:2005ad}
Amendola L.,  Barbieri R.,  2006, \mn@doi [Phys. Lett. B] {10.1016/j.physletb.2006.08.069}, 642, 192

\bibitem[\protect\citeauthoryear{{Amon} \& {Efstathiou}}{{Amon} \& {Efstathiou}}{2022}]{AAGPE2022}
{Amon} A.,  {Efstathiou} G.,  2022, \mn@doi [\mnras] {10.1093/mnras/stac2429}, \href {https://ui.adsabs.harvard.edu/abs/2022MNRAS.516.5355A} {516, 5355}

\bibitem[\protect\citeauthoryear{{Amon} et~al.,}{{Amon} et~al.}{2022}]{Amon:2021}
{Amon} A.,  et~al., 2022, \mn@doi [\prd] {10.1103/PhysRevD.105.023514}, \href {https://ui.adsabs.harvard.edu/abs/2022PhRvD.105b3514A} {105, 023514}

\bibitem[\protect\citeauthoryear{{Amon} et~al.,}{{Amon} et~al.}{2023}]{amon:2022}
{Amon} A.,  et~al., 2023, \mn@doi [\mnras] {10.1093/mnras/stac2938}, \href {https://ui.adsabs.harvard.edu/abs/2023MNRAS.518..477A} {518, 477}

\bibitem[\protect\citeauthoryear{{Angulo}, {Zennaro}, {Contreras}, {Aric{\`o}}, {Pellejero-Iba{\~n}ez}  \& {St{\"u}cker}}{{Angulo} et~al.}{2021}]{Bacco}
{Angulo} R.~E.,  {Zennaro} M.,  {Contreras} S.,  {Aric{\`o}} G.,  {Pellejero-Iba{\~n}ez} M.,   {St{\"u}cker} J.,  2021, \mn@doi [\mnras] {10.1093/mnras/stab2018}, \href {https://ui.adsabs.harvard.edu/abs/2021MNRAS.507.5869A} {507, 5869}

\bibitem[\protect\citeauthoryear{{Antypas} et~al.,}{{Antypas} et~al.}{2022}]{Antypas_2022}
{Antypas} D.,  et~al., 2022, \mn@doi [arXiv e-prints] {10.48550/arXiv.2203.14915}, \href {https://ui.adsabs.harvard.edu/abs/2022arXiv220314915A} {p. arXiv:2203.14915}

\bibitem[\protect\citeauthoryear{{Arun}, {Gudennavar}  \& {Sivaram}}{{Arun} et~al.}{2017}]{DM_DE_review}
{Arun} K.,  {Gudennavar} S.~B.,   {Sivaram} C.,  2017, \mn@doi [Advances in Space Research] {10.1016/j.asr.2017.03.043}, \href {https://ui.adsabs.harvard.edu/abs/2017AdSpR..60..166A} {60, 166}

\bibitem[\protect\citeauthoryear{{Arvanitaki}, {Dimopoulos}, {Dubovsky}, {Kaloper}  \& {March-Russell}}{{Arvanitaki} et~al.}{2010}]{Arvanitaki_2010}
{Arvanitaki} A.,  {Dimopoulos} S.,  {Dubovsky} S.,  {Kaloper} N.,   {March-Russell} J.,  2010, \mn@doi [\prd] {10.1103/PhysRevD.81.123530}, \href {https://ui.adsabs.harvard.edu/abs/2010PhRvD..81l3530A} {81, 123530}

\bibitem[\protect\citeauthoryear{{Asgari} et~al.,}{{Asgari} et~al.}{2021}]{Asgari:2021}
{Asgari} M.,  et~al., 2021, \mn@doi [\aap] {10.1051/0004-6361/202039070}, \href {https://ui.adsabs.harvard.edu/abs/2021A&A...645A.104A} {645, A104}

\bibitem[\protect\citeauthoryear{Banik, Bovy, Bertone, Erkal  \& de Boer}{Banik et~al.}{2021}]{Banik:2019smi}
Banik N.,  Bovy J.,  Bertone G.,  Erkal D.,   de Boer T. J.~L.,  2021, \mn@doi [JCAP] {10.1088/1475-7516/2021/10/043}, 10, 043

\bibitem[\protect\citeauthoryear{{Bauer}, {Marsh}, {Hlo{\v{z}}ek}, {Padmanabhan}  \& {Lagu{\"e}}}{{Bauer} et~al.}{2021}]{Bauer2021}
{Bauer} J.~B.,  {Marsh} D. J.~E.,  {Hlo{\v{z}}ek} R.,  {Padmanabhan} H.,   {Lagu{\"e}} A.,  2021, \mn@doi [\mnras] {10.1093/mnras/staa3300}, \href {https://ui.adsabs.harvard.edu/abs/2021MNRAS.500.3162B} {500, 3162}

\bibitem[\protect\citeauthoryear{Bigwood et~al.}{Bigwood et~al.}{2024}]{Bigwood:2024}
Bigwood L.,  et~al., 2024, \mn@doi [Mon. Not. Roy. Astron. Soc.] {10.1093/mnras/stae2100}, 534, 655

\bibitem[\protect\citeauthoryear{{Bocquet} et~al.,}{{Bocquet} et~al.}{2025}]{SPT_cosmology}
{Bocquet} S.,  et~al., 2025, \mn@doi [\prd] {10.1103/PhysRevD.111.063533}, \href {https://ui.adsabs.harvard.edu/abs/2025PhRvD.111f3533B} {111, 063533}

\bibitem[\protect\citeauthoryear{Bond, Cole, Efstathiou  \& Kaiser}{Bond et~al.}{1991}]{Excursion_set}
Bond J.~R.,  Cole S.,  Efstathiou G.,   Kaiser N.,  1991, \mn@doi [Astrophys. J.] {10.1086/170520}, 379, 440

\bibitem[\protect\citeauthoryear{{Bozek}, {Marsh}, {Silk}  \& {Wyse}}{{Bozek} et~al.}{2015}]{Bozek}
{Bozek} B.,  {Marsh} D. J.~E.,  {Silk} J.,   {Wyse} R. F.~G.,  2015, \mn@doi [\mnras] {10.1093/mnras/stv624}, \href {https://ui.adsabs.harvard.edu/abs/2015MNRAS.450..209B} {450, 209}

\bibitem[\protect\citeauthoryear{{Bridle} \& {King}}{{Bridle} \& {King}}{2007}]{Bridle_KingNLA}
{Bridle} S.,  {King} L.,  2007, \mn@doi [New Journal of Physics] {10.1088/1367-2630/9/12/444}, \href {https://ui.adsabs.harvard.edu/abs/2007NJPh....9..444B} {9, 444}

\bibitem[\protect\citeauthoryear{{Broxterman} \& {Kuijken}}{{Broxterman} \& {Kuijken}}{2024}]{Broxterman_Kuijken}
{Broxterman} J.~C.,  {Kuijken} K.,  2024, \mn@doi [\aap] {10.1051/0004-6361/202452319}, \href {https://ui.adsabs.harvard.edu/abs/2024A&A...692A.201B} {692, A201}

\bibitem[\protect\citeauthoryear{{Bullock} \& {Boylan-Kolchin}}{{Bullock} \& {Boylan-Kolchin}}{2017}]{Bullock_smallscale}
{Bullock} J.~S.,  {Boylan-Kolchin} M.,  2017, \mn@doi [\araa] {10.1146/annurev-astro-091916-055313}, \href {https://ui.adsabs.harvard.edu/abs/2017ARA&A..55..343B} {55, 343}

\bibitem[\protect\citeauthoryear{{Castorina}, {Sefusatti}, {Sheth}, {Villaescusa-Navarro}  \& {Viel}}{{Castorina} et~al.}{2014}]{Villaescusa-Navarro_2}
{Castorina} E.,  {Sefusatti} E.,  {Sheth} R.~K.,  {Villaescusa-Navarro} F.,   {Viel} M.,  2014, \mn@doi [\jcap] {10.1088/1475-7516/2014/02/049}, \href {https://ui.adsabs.harvard.edu/abs/2014JCAP...02..049C} {2014, 049}

\bibitem[\protect\citeauthoryear{{Chabanier} et~al.,}{{Chabanier} et~al.}{2019}]{Chabanier:2019b}
{Chabanier} S.,  et~al., 2019, \mn@doi [\jcap] {10.1088/1475-7516/2019/07/017}, \href {https://ui.adsabs.harvard.edu/abs/2019JCAP...07..017C} {2019, 017}

\bibitem[\protect\citeauthoryear{{Chisari} et~al.,}{{Chisari} et~al.}{2019}]{Chisari_2019}
{Chisari} N.~E.,  et~al., 2019, \mn@doi [The Open Journal of Astrophysics] {10.21105/astro.1905.06082}, \href {https://ui.adsabs.harvard.edu/abs/2019OJAp....2E...4C} {2, 4}

\bibitem[\protect\citeauthoryear{{Costanzi}, {Villaescusa-Navarro}, {Viel}, {Xia}, {Borgani}, {Castorina}  \& {Sefusatti}}{{Costanzi} et~al.}{2013}]{Villaescusa-Navarro_3}
{Costanzi} M.,  {Villaescusa-Navarro} F.,  {Viel} M.,  {Xia} J.-Q.,  {Borgani} S.,  {Castorina} E.,   {Sefusatti} E.,  2013, \mn@doi [\jcap] {10.1088/1475-7516/2013/12/012}, \href {https://ui.adsabs.harvard.edu/abs/2013JCAP...12..012C} {2013, 012}

\bibitem[\protect\citeauthoryear{{DES} \& {KiDS Collaborations}}{{DES} \& {KiDS Collaborations}}{2023}]{KiDSDES}
{DES} {KiDS Collaborations} 2023, The Open Journal of Astrophysics

\bibitem[\protect\citeauthoryear{{DESI Collaboration} et~al.,}{{DESI Collaboration} et~al.}{2025}]{DESI_BAO_DR2}
{DESI Collaboration} et~al., 2025, \mn@doi [arXiv e-prints] {10.48550/arXiv.2503.14738}, \href {https://ui.adsabs.harvard.edu/abs/2025arXiv250314738D} {p. arXiv:2503.14738}

\bibitem[\protect\citeauthoryear{{Dalal} \& {Kravtsov}}{{Dalal} \& {Kravtsov}}{2022}]{Dalal_2022}
{Dalal} N.,  {Kravtsov} A.,  2022, \mn@doi [\prd] {10.1103/PhysRevD.106.063517}, \href {https://ui.adsabs.harvard.edu/abs/2022PhRvD.106f3517D} {106, 063517}

\bibitem[\protect\citeauthoryear{Dalal et~al.,}{Dalal et~al.}{2023}]{dalal2023hyper}
Dalal R.,  et~al., 2023, Hyper Suprime-Cam Year 3 Results: Cosmology from Cosmic Shear Power Spectra (\mn@eprint {arXiv} {2304.00701})

\bibitem[\protect\citeauthoryear{{Dentler}, {Marsh}, {Hlo{\v{z}}ek}, {Lagu{\"e}}, {Rogers}  \& {Grin}}{{Dentler} et~al.}{2022}]{Dentler_Axions_weaklensing}
{Dentler} M.,  {Marsh} D. J.~E.,  {Hlo{\v{z}}ek} R.,  {Lagu{\"e}} A.,  {Rogers} K.~K.,   {Grin} D.,  2022, \mn@doi [\mnras] {10.1093/mnras/stac1946}, \href {https://ui.adsabs.harvard.edu/abs/2022MNRAS.515.5646D} {515, 5646}

\bibitem[\protect\citeauthoryear{Dine \& Fischler}{Dine \& Fischler}{1983}]{Dine_Fischler}
Dine M.,  Fischler W.,  1983, \mn@doi [Phys. Lett. B] {10.1016/0370-2693(83)90639-1}, 120, 137

\bibitem[\protect\citeauthoryear{{Dome}, {May}, {Lagu{\"e}}, {Marsh}, {Johnston}, {Bose}, {Tocher}  \& {Fialkov}}{{Dome} et~al.}{2025}]{Dome_Tibor}
{Dome} T.,  {May} S.,  {Lagu{\"e}} A.,  {Marsh} D. J.~E.,  {Johnston} S.,  {Bose} S.,  {Tocher} A.,   {Fialkov} A.,  2025, \mn@doi [\mnras] {10.1093/mnras/staf005}, \href {https://ui.adsabs.harvard.edu/abs/2025MNRAS.537..252D} {537, 252}

\bibitem[\protect\citeauthoryear{{Doux} et~al.,}{{Doux} et~al.}{2022}]{DouxDESHarmonic}
{Doux} C.,  et~al., 2022, \mn@doi [\mnras] {10.1093/mnras/stac1826}, \href {https://ui.adsabs.harvard.edu/abs/2022MNRAS.515.1942D} {515, 1942}

\bibitem[\protect\citeauthoryear{{Drlica-Wagner} et~al.,}{{Drlica-Wagner} et~al.}{2019}]{LSST_DM_summary}
{Drlica-Wagner} A.,  et~al., 2019, \mn@doi [arXiv e-prints] {10.48550/arXiv.1902.01055}, \href {https://ui.adsabs.harvard.edu/abs/2019arXiv190201055D} {p. arXiv:1902.01055}

\bibitem[\protect\citeauthoryear{Dvorkin et~al.}{Dvorkin et~al.}{2022}]{Dvorkin:2022bsc}
Dvorkin C.,  et~al., 2022, in {Snowmass 2021}.  (\mn@eprint {arXiv} {2203.07064})

\bibitem[\protect\citeauthoryear{Efstathiou}{Efstathiou}{2025}]{Efstathiou_RS}
Efstathiou G.,  2025, \mn@doi [Phil. Trans. Roy. Soc. Lond. A] {10.1098/rsta.2024.0022}, 383, 20240022

\bibitem[\protect\citeauthoryear{{Efstathiou} \& {Gratton}}{{Efstathiou} \& {Gratton}}{2020}]{Efstathiou:2020}
{Efstathiou} G.,  {Gratton} S.,  2020, \mn@doi [\mnras] {10.1093/mnrasl/slaa093}, \href {https://ui.adsabs.harvard.edu/abs/2020MNRAS.496L..91E} {496, L91}

\bibitem[\protect\citeauthoryear{{Efstathiou} \& {McCarthy}}{{Efstathiou} \& {McCarthy}}{2025}]{Efstathiou:2025}
{Efstathiou} G.,  {McCarthy} F.,  2025, \mn@doi [arXiv e-prints] {10.48550/arXiv.2502.10232}, \href {https://ui.adsabs.harvard.edu/abs/2025arXiv250210232E} {p. arXiv:2502.10232}

\bibitem[\protect\citeauthoryear{{Efstathiou}, {Rosenberg}  \& {Poulin}}{{Efstathiou} et~al.}{2024}]{Efstathiou_EDE}
{Efstathiou} G.,  {Rosenberg} E.,   {Poulin} V.,  2024, \mn@doi [\prl] {10.1103/PhysRevLett.132.221002}, \href {https://ui.adsabs.harvard.edu/abs/2024PhRvL.132v1002E} {132, 221002}

\bibitem[\protect\citeauthoryear{{Euclid Collaboration} et~al.,}{{Euclid Collaboration} et~al.}{2020}]{EuclidForecast}
{Euclid Collaboration} et~al., 2020, \mn@doi [\aap] {10.1051/0004-6361/202038071}, \href {https://ui.adsabs.harvard.edu/abs/2020A&A...642A.191E} {642, A191}

\bibitem[\protect\citeauthoryear{{Euclid Collaboration} et~al.,}{{Euclid Collaboration} et~al.}{2021}]{EuclidEmulator}
{Euclid Collaboration} et~al., 2021, \mn@doi [\mnras] {10.1093/mnras/stab1366}, \href {https://ui.adsabs.harvard.edu/abs/2021MNRAS.505.2840E} {505, 2840}

\bibitem[\protect\citeauthoryear{{Euclid Collaboration} et~al.,}{{Euclid Collaboration} et~al.}{2024}]{Koyama_Euclid}
{Euclid Collaboration} et~al., 2024, \mn@doi [arXiv e-prints] {10.48550/arXiv.2409.03524}, \href {https://ui.adsabs.harvard.edu/abs/2024arXiv240903524E} {p. arXiv:2409.03524}

\bibitem[\protect\citeauthoryear{{Farren}, {Grin}, {Jaffe}, {Hlo{\v{z}}ek}  \& {Marsh}}{{Farren} et~al.}{2022}]{Farren_axions}
{Farren} G.~S.,  {Grin} D.,  {Jaffe} A.~H.,  {Hlo{\v{z}}ek} R.,   {Marsh} D. J.~E.,  2022, \mn@doi [\prd] {10.1103/PhysRevD.105.063513}, \href {https://ui.adsabs.harvard.edu/abs/2022PhRvD.105f3513F} {105, 063513}

\bibitem[\protect\citeauthoryear{Feroz, Hobson  \& Bridges}{Feroz et~al.}{2009}]{Feroz_2009}
Feroz F.,  Hobson M.~P.,   Bridges M.,  2009, \mn@doi [Monthly Notices of the Royal Astronomical Society] {10.1111/j.1365-2966.2009.14548.x}, 398, 1601

\bibitem[\protect\citeauthoryear{Ferreira, Alonso, Garcia-Garcia  \& Chisari}{Ferreira et~al.}{2024}]{Ferreira:2023syi}
Ferreira T.,  Alonso D.,  Garcia-Garcia C.,   Chisari N.~E.,  2024, \mn@doi [Phys. Rev. Lett.] {10.1103/PhysRevLett.133.051001}, 133, 051001

\bibitem[\protect\citeauthoryear{{Giri} \& {Schneider}}{{Giri} \& {Schneider}}{2023}]{BCEmu}
{Giri} S.~K.,  {Schneider} A.,  2023, {BCemu: Model baryonic effects in cosmological simulations}, Astrophysics Source Code Library, record ascl:2308.010

\bibitem[\protect\citeauthoryear{{Green}}{{Green}}{2024}]{Green_PBH_review}
{Green} A.~M.,  2024, \mn@doi [Nuclear Physics B] {10.1016/j.nuclphysb.2024.116494}, \href {https://ui.adsabs.harvard.edu/abs/2024NuPhB100316494G} {1003, 116494}

\bibitem[\protect\citeauthoryear{{Grin}, {Marsh}  \& {Hlozek}}{{Grin} et~al.}{2022}]{axioncamb}
{Grin} D.,  {Marsh} D. J.~E.,   {Hlozek} R.,  2022, {axionCAMB: Modification of the CAMB Boltzmann code}, Astrophysics Source Code Library, record ascl:2203.026

\bibitem[\protect\citeauthoryear{Heitmann, Higdon, White, Habib, Williams  \& Wagner}{Heitmann et~al.}{2009}]{Heitmann:2009cu}
Heitmann K.,  Higdon D.,  White M.,  Habib S.,  Williams B.~J.,   Wagner C.,  2009, \mn@doi [Astrophys. J.] {10.1088/0004-637X/705/1/156}, 705, 156

\bibitem[\protect\citeauthoryear{{Heymans} et~al.,}{{Heymans} et~al.}{2013}]{Heymans:2013}
{Heymans} C.,  et~al., 2013, \mn@doi [\mnras] {10.1093/mnras/stt601}, \href {http://adsabs.harvard.edu/abs/2013MNRAS.432.2433H} {432, 2433}

\bibitem[\protect\citeauthoryear{{Heymans} et~al.,}{{Heymans} et~al.}{2021}]{Heymans:2021}
{Heymans} C.,  et~al., 2021, \mn@doi [\aap] {10.1051/0004-6361/202039063}, \href {https://ui.adsabs.harvard.edu/abs/2021A&A...646A.140H} {646, A140}

\bibitem[\protect\citeauthoryear{{Hildebrandt} et~al.,}{{Hildebrandt} et~al.}{2017}]{Hildebrandt:2017}
{Hildebrandt} H.,  et~al., 2017, \mn@doi [\mnras] {10.1093/mnras/stw2805}, \href {https://ui.adsabs.harvard.edu/abs/2017MNRAS.465.1454H} {465, 1454}

\bibitem[\protect\citeauthoryear{{Hlo{\v{z}}ek}, {Marsh}  \& {Grin}}{{Hlo{\v{z}}ek} et~al.}{2018}]{Hlozek_CMB_axions2018}
{Hlo{\v{z}}ek} R.,  {Marsh} D. J.~E.,   {Grin} D.,  2018, \mn@doi [\mnras] {10.1093/mnras/sty271}, \href {https://ui.adsabs.harvard.edu/abs/2018MNRAS.476.3063H} {476, 3063}

\bibitem[\protect\citeauthoryear{{Hlozek}, {Grin}, {Marsh}  \& {Ferreira}}{{Hlozek} et~al.}{2015}]{Hlozek_CMB_axions2015}
{Hlozek} R.,  {Grin} D.,  {Marsh} D. J.~E.,   {Ferreira} P.~G.,  2015, \mn@doi [\prd] {10.1103/PhysRevD.91.103512}, \href {https://ui.adsabs.harvard.edu/abs/2015PhRvD..91j3512H} {91, 103512}

\bibitem[\protect\citeauthoryear{{Hotinli}, {Marsh}  \& {Kamionkowski}}{{Hotinli} et~al.}{2022}]{Hotinli_2022}
{Hotinli} S.~C.,  {Marsh} D. J.~E.,   {Kamionkowski} M.,  2022, \mn@doi [\prd] {10.1103/PhysRevD.106.043529}, \href {https://ui.adsabs.harvard.edu/abs/2022PhRvD.106d3529H} {106, 043529}

\bibitem[\protect\citeauthoryear{{Hu}, {Barkana}  \& {Gruzinov}}{{Hu} et~al.}{2000}]{Hu:2000}
{Hu} W.,  {Barkana} R.,   {Gruzinov} A.,  2000, \mn@doi [\prl] {10.1103/PhysRevLett.85.1158}, \href {https://ui.adsabs.harvard.edu/abs/2000PhRvL..85.1158H} {85, 1158}

\bibitem[\protect\citeauthoryear{{Hui}, {Ostriker}, {Tremaine}  \& {Witten}}{{Hui} et~al.}{2017}]{Hui:2017}
{Hui} L.,  {Ostriker} J.~P.,  {Tremaine} S.,   {Witten} E.,  2017, \mn@doi [\prd] {10.1103/PhysRevD.95.043541}, \href {https://ui.adsabs.harvard.edu/abs/2017PhRvD..95d3541H} {95, 043541}

\bibitem[\protect\citeauthoryear{{Hwang} \& {Noh}}{{Hwang} \& {Noh}}{2009}]{Hwang_axion}
{Hwang} J.-C.,  {Noh} H.,  2009, \mn@doi [Physics Letters B] {10.1016/j.physletb.2009.08.031}, \href {https://ui.adsabs.harvard.edu/abs/2009PhLB..680....1H} {680, 1}

\bibitem[\protect\citeauthoryear{{Ir{\v{s}}i{\v{c}}}, {Viel}, {Haehnelt}, {Bolton}  \& {Becker}}{{Ir{\v{s}}i{\v{c}}} et~al.}{2017}]{Irsic_Lyman_DM}
{Ir{\v{s}}i{\v{c}}} V.,  {Viel} M.,  {Haehnelt} M.~G.,  {Bolton} J.~S.,   {Becker} G.~D.,  2017, \mn@doi [\prl] {10.1103/PhysRevLett.119.031302}, \href {https://ui.adsabs.harvard.edu/abs/2017PhRvL.119c1302I} {119, 031302}

\bibitem[\protect\citeauthoryear{{Kadan}}{{Kadan}}{2024}]{Kadan:2024}
{Kadan} S.,  2024, \mn@doi [arXiv e-prints] {10.48550/arXiv.2404.16922}, \href {https://ui.adsabs.harvard.edu/abs/2024arXiv240416922K} {p. arXiv:2404.16922}

\bibitem[\protect\citeauthoryear{Khlopov, Malomed, Zeldovich  \& Zeldovich}{Khlopov et~al.}{1985}]{Khlopov_1985}
Khlopov M.~Y.,  Malomed B.~A.,  Zeldovich I.~B.,   Zeldovich Y.~B.,  1985, \mn@doi [Mon. Not. Roy. Astron. Soc.] {10.1093/mnras/215.4.575}, 215, 575

\bibitem[\protect\citeauthoryear{Khmelnitsky \& Rubakov}{Khmelnitsky \& Rubakov}{2014}]{Khmelnitsky:2013lxt}
Khmelnitsky A.,  Rubakov V.,  2014, \mn@doi [JCAP] {10.1088/1475-7516/2014/02/019}, 02, 019

\bibitem[\protect\citeauthoryear{{Kobayashi}, {Murgia}, {De Simone}, {Ir{\v{s}}i{\v{c}}}  \& {Viel}}{{Kobayashi} et~al.}{2017}]{Kobayashi_Lyman_DM}
{Kobayashi} T.,  {Murgia} R.,  {De Simone} A.,  {Ir{\v{s}}i{\v{c}}} V.,   {Viel} M.,  2017, \mn@doi [\prd] {10.1103/PhysRevD.96.123514}, \href {https://ui.adsabs.harvard.edu/abs/2017PhRvD..96l3514K} {96, 123514}

\bibitem[\protect\citeauthoryear{{Kunkel}, {Chiueh}  \& {Sch{\"a}fer}}{{Kunkel} et~al.}{2022}]{Kunkel_FDM_lensing}
{Kunkel} A.,  {Chiueh} T.,   {Sch{\"a}fer} B.~M.,  2022, \mn@doi [arXiv e-prints] {10.48550/arXiv.2211.01523}, \href {https://ui.adsabs.harvard.edu/abs/2022arXiv221101523K} {p. arXiv:2211.01523}

\bibitem[\protect\citeauthoryear{{LSST Science Collaboration} et~al.,}{{LSST Science Collaboration} et~al.}{2009}]{LSSTScience}
{LSST Science Collaboration} et~al., 2009, \mn@doi [arXiv e-prints] {10.48550/arXiv.0912.0201}, \href {https://ui.adsabs.harvard.edu/abs/2009arXiv0912.0201L} {p. arXiv:0912.0201}

\bibitem[\protect\citeauthoryear{{Lagu{\"e}}, {Bond}, {Hlo{\v{z}}ek}, {Rogers}, {Marsh}  \& {Grin}}{{Lagu{\"e}} et~al.}{2022}]{Lague_axions_galaxyclustering}
{Lagu{\"e}} A.,  {Bond} J.~R.,  {Hlo{\v{z}}ek} R.,  {Rogers} K.~K.,  {Marsh} D.~J.~E.,   {Grin} D.,  2022, \mn@doi [\jcap] {10.1088/1475-7516/2022/01/049}, \href {https://ui.adsabs.harvard.edu/abs/2022JCAP...01..049L} {2022, 049}

\bibitem[\protect\citeauthoryear{{Lagu{\"e}}, {Schwabe}, {Hlo{\v{z}}ek}, {Marsh}  \& {Rogers}}{{Lagu{\"e}} et~al.}{2024}]{Lague}
{Lagu{\"e}} A.,  {Schwabe} B.,  {Hlo{\v{z}}ek} R.,  {Marsh} D. J.~E.,   {Rogers} K.~K.,  2024, \mn@doi [\prd] {10.1103/PhysRevD.109.043507}, \href {https://ui.adsabs.harvard.edu/abs/2024PhRvD.109d3507L} {109, 043507}

\bibitem[\protect\citeauthoryear{{Lamman}, {Tsaprazi}, {Shi}, {{\v{S}}ar{\v{c}}evi{\'c}}, {Pyne}, {Legnani}  \& {Ferreira}}{{Lamman} et~al.}{2024}]{Lamman:2024}
{Lamman} C.,  {Tsaprazi} E.,  {Shi} J.,  {{\v{S}}ar{\v{c}}evi{\'c}} N.~N.,  {Pyne} S.,  {Legnani} E.,   {Ferreira} T.,  2024, \mn@doi [The Open Journal of Astrophysics] {10.21105/astro.2309.08605}, \href {https://ui.adsabs.harvard.edu/abs/2024OJAp....7E..14L} {7, 14}

\bibitem[\protect\citeauthoryear{{Laroche}, {Gilman}, {Li}, {Bovy}  \& {Du}}{{Laroche} et~al.}{2022}]{Laroche_etal}
{Laroche} A.,  {Gilman} D.,  {Li} X.,  {Bovy} J.,   {Du} X.,  2022, \mn@doi [\mnras] {10.1093/mnras/stac2677}, \href {https://ui.adsabs.harvard.edu/abs/2022MNRAS.517.1867L} {517, 1867}

\bibitem[\protect\citeauthoryear{{Le Brun}, {McCarthy}, {Schaye}  \& {Ponman}}{{Le Brun} et~al.}{2014}]{COSMOowls}
{Le Brun} A. M.~C.,  {McCarthy} I.~G.,  {Schaye} J.,   {Ponman} T.~J.,  2014, \mn@doi [\mnras] {10.1093/mnras/stu608}, \href {https://ui.adsabs.harvard.edu/abs/2014MNRAS.441.1270L} {441, 1270}

\bibitem[\protect\citeauthoryear{Li et~al.,}{Li et~al.}{2023}]{li2023hyper}
Li X.,  et~al., 2023, Hyper Suprime-Cam Year 3 Results: Cosmology from Cosmic Shear Two-point Correlation Functions (\mn@eprint {arXiv} {2304.00702})

\bibitem[\protect\citeauthoryear{{Liu}, {Hu}  \& {Grin}}{{Liu} et~al.}{2024}]{Liu_axions}
{Liu} R.,  {Hu} W.,   {Grin} D.,  2024, \mn@doi [arXiv e-prints] {10.48550/arXiv.2412.15192}, \href {https://ui.adsabs.harvard.edu/abs/2024arXiv241215192L} {p. arXiv:2412.15192}

\bibitem[\protect\citeauthoryear{{Louis} et~al.,}{{Louis} et~al.}{2025}]{ACT_primary_CMB}
{Louis} T.,  et~al., 2025, \mn@doi [arXiv e-prints] {10.48550/arXiv.2503.14452}, \href {https://ui.adsabs.harvard.edu/abs/2025arXiv250314452L} {p. arXiv:2503.14452}

\bibitem[\protect\citeauthoryear{{Marsh}}{{Marsh}}{2016}]{Marsh_axion}
{Marsh} D. J.~E.,  2016, \mn@doi [\physrep] {10.1016/j.physrep.2016.06.005}, \href {https://ui.adsabs.harvard.edu/abs/2016PhR...643....1M} {643, 1}

\bibitem[\protect\citeauthoryear{{Marsh} \& {Ferreira}}{{Marsh} \& {Ferreira}}{2010}]{Marsh_Ferreira_2010}
{Marsh} D. J.~E.,  {Ferreira} P.~G.,  2010, \mn@doi [\prd] {10.1103/PhysRevD.82.103528}, \href {https://ui.adsabs.harvard.edu/abs/2010PhRvD..82j3528M} {82, 103528}

\bibitem[\protect\citeauthoryear{{Marsh} \& {Niemeyer}}{{Marsh} \& {Niemeyer}}{2019}]{Marsh_Niemeyer}
{Marsh} D. J.~E.,  {Niemeyer} J.~C.,  2019, \mn@doi [\prl] {10.1103/PhysRevLett.123.051103}, \href {https://ui.adsabs.harvard.edu/abs/2019PhRvL.123e1103M} {123, 051103}

\bibitem[\protect\citeauthoryear{{Marsh} \& {Pop}}{{Marsh} \& {Pop}}{2015}]{Marsh_cuspcore}
{Marsh} D. J.~E.,  {Pop} A.-R.,  2015, \mn@doi [\mnras] {10.1093/mnras/stv1050}, \href {https://ui.adsabs.harvard.edu/abs/2015MNRAS.451.2479M} {451, 2479}

\bibitem[\protect\citeauthoryear{{Marsh} \& {Silk}}{{Marsh} \& {Silk}}{2014}]{Marsh_Silk}
{Marsh} D. J.~E.,  {Silk} J.,  2014, \mn@doi [\mnras] {10.1093/mnras/stt2079}, \href {https://ui.adsabs.harvard.edu/abs/2014MNRAS.437.2652M} {437, 2652}

\bibitem[\protect\citeauthoryear{{Massara}, {Villaescusa-Navarro}  \& {Viel}}{{Massara} et~al.}{2014}]{Massara2024}
{Massara} E.,  {Villaescusa-Navarro} F.,   {Viel} M.,  2014, \mn@doi [\jcap] {10.1088/1475-7516/2014/12/053}, \href {https://ui.adsabs.harvard.edu/abs/2014JCAP...12..053M} {2014, 053}

\bibitem[\protect\citeauthoryear{{May} \& {Springel}}{{May} \& {Springel}}{2021}]{May_Springel_FDM}
{May} S.,  {Springel} V.,  2021, \mn@doi [\mnras] {10.1093/mnras/stab1764}, \href {https://ui.adsabs.harvard.edu/abs/2021MNRAS.506.2603M} {506, 2603}

\bibitem[\protect\citeauthoryear{{McCarthy}, {Schaye}, {Bird}  \& {Le Brun}}{{McCarthy} et~al.}{2017}]{BAHAMAS}
{McCarthy} I.~G.,  {Schaye} J.,  {Bird} S.,   {Le Brun} A. M.~C.,  2017, \mn@doi [\mnras] {10.1093/mnras/stw2792}, \href {https://ui.adsabs.harvard.edu/abs/2017MNRAS.465.2936M} {465, 2936}

\bibitem[\protect\citeauthoryear{{McCullough} et~al.,}{{McCullough} et~al.}{2024}]{McCullough_BlueIA}
{McCullough} J.,  et~al., 2024, \mn@doi [arXiv e-prints] {10.48550/arXiv.2410.22272}, \href {https://ui.adsabs.harvard.edu/abs/2024arXiv241022272M} {p. arXiv:2410.22272}

\bibitem[\protect\citeauthoryear{{Mead}, {Brieden}, {Tr{\"o}ster}  \& {Heymans}}{{Mead} et~al.}{2021}]{mead:2021}
{Mead} A.~J.,  {Brieden} S.,  {Tr{\"o}ster} T.,   {Heymans} C.,  2021, \mn@doi [\mnras] {10.1093/mnras/stab082}, \href {https://ui.adsabs.harvard.edu/abs/2021MNRAS.502.1401M} {502, 1401}

\bibitem[\protect\citeauthoryear{{Mocz} \& {Succi}}{{Mocz} \& {Succi}}{2015}]{Mocz_2015}
{Mocz} P.,  {Succi} S.,  2015, \mn@doi [\pre] {10.1103/PhysRevE.91.053304}, \href {https://ui.adsabs.harvard.edu/abs/2015PhRvE..91e3304M} {91, 053304}

\bibitem[\protect\citeauthoryear{{Mocz} et~al.,}{{Mocz} et~al.}{2019}]{Mocz_HYDROAXION}
{Mocz} P.,  et~al., 2019, \mn@doi [\prl] {10.1103/PhysRevLett.123.141301}, \href {https://ui.adsabs.harvard.edu/abs/2019PhRvL.123n1301M} {123, 141301}

\bibitem[\protect\citeauthoryear{{Nadler} et~al.,}{{Nadler} et~al.}{2021}]{2021PhRvL.126i1101N}
{Nadler} E.~O.,  et~al., 2021, \mn@doi [\prl] {10.1103/PhysRevLett.126.091101}, \href {https://ui.adsabs.harvard.edu/abs/2021PhRvL.126i1101N} {126, 091101}

\bibitem[\protect\citeauthoryear{{Navarro}, {Frenk}  \& {White}}{{Navarro} et~al.}{1997}]{NFW}
{Navarro} J.~F.,  {Frenk} C.~S.,   {White} S. D.~M.,  1997, \mn@doi [\apj] {10.1086/304888}, \href {https://ui.adsabs.harvard.edu/abs/1997ApJ...490..493N} {490, 493}

\bibitem[\protect\citeauthoryear{{Padmanabhan}, {Refregier}  \& {Amara}}{{Padmanabhan} et~al.}{2017}]{Padmanabhan_tracer}
{Padmanabhan} H.,  {Refregier} A.,   {Amara} A.,  2017, \mn@doi [\mnras] {10.1093/mnras/stx979}, \href {https://ui.adsabs.harvard.edu/abs/2017MNRAS.469.2323P} {469, 2323}

\bibitem[\protect\citeauthoryear{{Paopiamsap}, {Porqueres}, {Alonso}, {Harnois-Deraps}  \& {Leonard}}{{Paopiamsap} et~al.}{2024}]{Paopiamsap_DESC_IA}
{Paopiamsap} A.,  {Porqueres} N.,  {Alonso} D.,  {Harnois-Deraps} J.,   {Leonard} C.~D.,  2024, \mn@doi [The Open Journal of Astrophysics] {10.33232/001c.117419}, \href {https://ui.adsabs.harvard.edu/abs/2024OJAp....7E..34P} {7, 34}

\bibitem[\protect\citeauthoryear{{Peacock} \& {Smith}}{{Peacock} \& {Smith}}{2000}]{Peacock_halomodel}
{Peacock} J.~A.,  {Smith} R.~E.,  2000, \mn@doi [\mnras] {10.1046/j.1365-8711.2000.03779.x}, \href {https://ui.adsabs.harvard.edu/abs/2000MNRAS.318.1144P} {318, 1144}

\bibitem[\protect\citeauthoryear{Peccei \& Quinn}{Peccei \& Quinn}{1977}]{Peccei_strongCPaxion}
Peccei R.~D.,  Quinn H.~R.,  1977, \mn@doi [Phys. Rev. Lett.] {10.1103/PhysRevLett.38.1440}, 38, 1440

\bibitem[\protect\citeauthoryear{{Perez Sarmiento}, {Lagu{\"e}}, {Madhavacheril}, {Jain}  \& {Sherwin}}{{Perez Sarmiento} et~al.}{2025}]{Sarmiento}
{Perez Sarmiento} K.,  {Lagu{\"e}} A.,  {Madhavacheril} M.,  {Jain} B.,   {Sherwin} B.,  2025, \mn@doi [arXiv e-prints] {10.48550/arXiv.2502.06687}, \href {https://ui.adsabs.harvard.edu/abs/2025arXiv250206687P} {p. arXiv:2502.06687}

\bibitem[\protect\citeauthoryear{{Planck Collaboration} et~al.,}{{Planck Collaboration} et~al.}{2020a}]{Planck_legacy}
{Planck Collaboration} et~al., 2020a, \mn@doi [\aap] {10.1051/0004-6361/201833880}, \href {https://ui.adsabs.harvard.edu/abs/2020A&A...641A...1P} {641, A1}

\bibitem[\protect\citeauthoryear{{Planck Collaboration} et~al.,}{{Planck Collaboration} et~al.}{2020b}]{Params:2018}
{Planck Collaboration} et~al., 2020b, \mn@doi [\aap] {10.1051/0004-6361/201833910}, \href {https://ui.adsabs.harvard.edu/abs/2020A&A...641A...6P} {641, A6}

\bibitem[\protect\citeauthoryear{Poulin, Smith, Grin, Karwal  \& Kamionkowski}{Poulin et~al.}{2018}]{Axiclass_1}
Poulin V.,  Smith T.~L.,  Grin D.,  Karwal T.,   Kamionkowski M.,  2018, \mn@doi [Phys. Rev. D] {10.1103/PhysRevD.98.083525}, 98, 083525

\bibitem[\protect\citeauthoryear{{Poulin}, {Smith}  \& {Karwal}}{{Poulin} et~al.}{2023}]{Poulin_EDE}
{Poulin} V.,  {Smith} T.~L.,   {Karwal} T.,  2023, \mn@doi [Physics of the Dark Universe] {10.1016/j.dark.2023.101348}, \href {https://ui.adsabs.harvard.edu/abs/2023PDU....4201348P} {42, 101348}

\bibitem[\protect\citeauthoryear{Preskill, Wise  \& Wilczek}{Preskill et~al.}{1983}]{Preskill_axion}
Preskill J.,  Wise M.~B.,   Wilczek F.,  1983, \mn@doi [Phys. Lett. B] {10.1016/0370-2693(83)90637-8}, 120, 127

\bibitem[\protect\citeauthoryear{{Press} \& {Schechter}}{{Press} \& {Schechter}}{1974}]{Press_Schechter}
{Press} W.~H.,  {Schechter} P.,  1974, \mn@doi [\apj] {10.1086/152650}, \href {https://ui.adsabs.harvard.edu/abs/1974ApJ...187..425P} {187, 425}

\bibitem[\protect\citeauthoryear{{Preston}, {Amon}  \& {Efstathiou}}{{Preston} et~al.}{2023}]{CPAAGPE}
{Preston} C.,  {Amon} A.,   {Efstathiou} G.,  2023, \mn@doi [\mnras] {10.1093/mnras/stad2573}, \href {https://ui.adsabs.harvard.edu/abs/2023MNRAS.525.5554P} {525, 5554}

\bibitem[\protect\citeauthoryear{{Preston}, {Amon}  \& {Efstathiou}}{{Preston} et~al.}{2024}]{Preston_2024}
{Preston} C.,  {Amon} A.,   {Efstathiou} G.,  2024, \mn@doi [\mnras] {10.1093/mnras/stae1848}, \href {https://ui.adsabs.harvard.edu/abs/2024MNRAS.533..621P} {533, 621}

\bibitem[\protect\citeauthoryear{{Reed}, {Gardner}, {Quinn}, {Stadel}, {Fardal}, {Lake}  \& {Governato}}{{Reed} et~al.}{2003}]{Reed_2003}
{Reed} D.,  {Gardner} J.,  {Quinn} T.,  {Stadel} J.,  {Fardal} M.,  {Lake} G.,   {Governato} F.,  2003, \mn@doi [\mnras] {10.1046/j.1365-2966.2003.07113.x}, \href {https://ui.adsabs.harvard.edu/abs/2003MNRAS.346..565R} {346, 565}

\bibitem[\protect\citeauthoryear{{Riess} et~al.,}{{Riess} et~al.}{2022}]{reiss2022}
{Riess} A.~G.,  et~al., 2022, \mn@doi [\apjl] {10.3847/2041-8213/ac5c5b}, \href {https://ui.adsabs.harvard.edu/abs/2022ApJ...934L...7R} {934, L7}

\bibitem[\protect\citeauthoryear{Rogers \& Peiris}{Rogers \& Peiris}{2021a}]{Rogers:2020cup}
Rogers K.~K.,  Peiris H.~V.,  2021a, \mn@doi [Phys. Rev. D] {10.1103/PhysRevD.103.043526}, 103, 043526

\bibitem[\protect\citeauthoryear{{Rogers} \& {Peiris}}{{Rogers} \& {Peiris}}{2021b}]{Rogers_Peiris_Lyman}
{Rogers} K.~K.,  {Peiris} H.~V.,  2021b, \mn@doi [\prl] {10.1103/PhysRevLett.126.071302}, \href {https://ui.adsabs.harvard.edu/abs/2021PhRvL.126g1302R} {126, 071302}

\bibitem[\protect\citeauthoryear{Rogers \& Poulin}{Rogers \& Poulin}{2025}]{Rogers_Poulin}
Rogers K.~K.,  Poulin V.,  2025, \mn@doi [Phys. Rev. Res.] {10.1103/PhysRevResearch.7.L012018}, 7, L012018

\bibitem[\protect\citeauthoryear{Rogers, Peiris, Pontzen, Bird, Verde  \& Font-Ribera}{Rogers et~al.}{2019}]{Rogers:2018smb}
Rogers K.~K.,  Peiris H.~V.,  Pontzen A.,  Bird S.,  Verde L.,   Font-Ribera A.,  2019, \mn@doi [JCAP] {10.1088/1475-7516/2019/02/031}, 02, 031

\bibitem[\protect\citeauthoryear{{Rogers}, {Hlo{\v{z}}ek}, {Lagu{\"e}}, {Ivanov}, {Philcox}, {Cabass}, {Akitsu}  \& {Marsh}}{{Rogers} et~al.}{2023}]{Rogers_S8tensionaxions}
{Rogers} K.~K.,  {Hlo{\v{z}}ek} R.,  {Lagu{\"e}} A.,  {Ivanov} M.~M.,  {Philcox} O. H.~E.,  {Cabass} G.,  {Akitsu} K.,   {Marsh} D. J.~E.,  2023, \mn@doi [\jcap] {10.1088/1475-7516/2023/06/023}, \href {https://ui.adsabs.harvard.edu/abs/2023JCAP...06..023R} {2023, 023}

\bibitem[\protect\citeauthoryear{{Roszkowski}, {Sessolo}  \& {Trojanowski}}{{Roszkowski} et~al.}{2018}]{Roszkowski_2018}
{Roszkowski} L.,  {Sessolo} E.~M.,   {Trojanowski} S.,  2018, \mn@doi [Reports on Progress in Physics] {10.1088/1361-6633/aab913}, \href {https://ui.adsabs.harvard.edu/abs/2018RPPh...81f6201R} {81, 066201}

\bibitem[\protect\citeauthoryear{{Rubin}, {Ford}  \& {Thonnard}}{{Rubin} et~al.}{1980}]{Rubin_DM}
{Rubin} V.~C.,  {Ford} W.~K. J.,   {Thonnard} N.,  1980, \mn@doi [\apj] {10.1086/158003}, \href {https://ui.adsabs.harvard.edu/abs/1980ApJ...238..471R} {238, 471}

\bibitem[\protect\citeauthoryear{{Schaye} et~al.,}{{Schaye} et~al.}{2023}]{flamingo}
{Schaye} J.,  et~al., 2023, \mn@doi [\mnras] {10.1093/mnras/stad2419}, \href {https://ui.adsabs.harvard.edu/abs/2023MNRAS.526.4978S} {526, 4978}

\bibitem[\protect\citeauthoryear{{Schive}, {Chiueh}  \& {Broadhurst}}{{Schive} et~al.}{2014}]{Schive_axiondensity}
{Schive} H.-Y.,  {Chiueh} T.,   {Broadhurst} T.,  2014, \mn@doi [Nature Physics] {10.1038/nphys2996}, \href {https://ui.adsabs.harvard.edu/abs/2014NatPh..10..496S} {10, 496}

\bibitem[\protect\citeauthoryear{{Schneider} et~al.,}{{Schneider} et~al.}{2020}]{Schneider:2020}
{Schneider} A.,  et~al., 2020, \mn@doi [\jcap] {10.1088/1475-7516/2020/04/020}, \href {https://ui.adsabs.harvard.edu/abs/2020JCAP...04..020S} {2020, 020}

\bibitem[\protect\citeauthoryear{{Schneider}, {Giri}, {Amodeo}  \& {Refregier}}{{Schneider} et~al.}{2021}]{schneider:2021}
{Schneider} A.,  {Giri} S.~K.,  {Amodeo} S.,   {Refregier} A.,  2021, arXiv e-prints, \href {https://ui.adsabs.harvard.edu/abs/2021arXiv211002228S} {p. arXiv:2110.02228}

\bibitem[\protect\citeauthoryear{Schwabe, Gosenca, Behrens, Niemeyer  \& Easther}{Schwabe et~al.}{2020}]{Schwabe_sims}
Schwabe B.,  Gosenca M.,  Behrens C.,  Niemeyer J.~C.,   Easther R.,  2020, \mn@doi [Phys. Rev. D] {10.1103/PhysRevD.102.083518}, 102, 083518

\bibitem[\protect\citeauthoryear{{Secco}, {Samuroff}  et~al.}{{Secco} et~al.}{2022}]{Secco:2022}
{Secco} L.~F.,  {Samuroff} S.,   et~al., 2022, \mn@doi [\prd] {10.1103/PhysRevD.105.023515}, \href {https://ui.adsabs.harvard.edu/abs/2022PhRvD.105b3515S} {105, 023515}

\bibitem[\protect\citeauthoryear{{Seljak}}{{Seljak}}{2000}]{Seljak_halomodel}
{Seljak} U.,  2000, \mn@doi [\mnras] {10.1046/j.1365-8711.2000.03715.x}, \href {https://ui.adsabs.harvard.edu/abs/2000MNRAS.318..203S} {318, 203}

\bibitem[\protect\citeauthoryear{Servant \& Simakachorn}{Servant \& Simakachorn}{2023}]{Servant_PTA_axions}
Servant G.,  Simakachorn P.,  2023, \mn@doi [Phys. Rev. D] {10.1103/PhysRevD.108.123516}, 108, 123516

\bibitem[\protect\citeauthoryear{{Sheridan} et~al.,}{{Sheridan} et~al.}{2024}]{Sheridan_2024}
{Sheridan} E.,  et~al., 2024, \mn@doi [arXiv e-prints] {10.48550/arXiv.2412.12012}, \href {https://ui.adsabs.harvard.edu/abs/2024arXiv241212012S} {p. arXiv:2412.12012}

\bibitem[\protect\citeauthoryear{{Sheth} \& {Tormen}}{{Sheth} \& {Tormen}}{2002}]{Sheth_Tormen}
{Sheth} R.~K.,  {Tormen} G.,  2002, \mn@doi [\mnras] {10.1046/j.1365-8711.2002.04950.x}, \href {https://ui.adsabs.harvard.edu/abs/2002MNRAS.329...61S} {329, 61}

\bibitem[\protect\citeauthoryear{{Siegel} et~al.,}{{Siegel} et~al.}{2025}]{Siegel_DESI_IA}
{Siegel} J.,  et~al., 2025, \mn@doi [arXiv e-prints] {10.48550/arXiv.2507.11530}, \href {https://ui.adsabs.harvard.edu/abs/2025arXiv250711530S} {p. arXiv:2507.11530}

\bibitem[\protect\citeauthoryear{Smarra et~al.}{Smarra et~al.}{2023}]{EuropeanPulsarTimingArray:2023egv}
Smarra C.,  et~al., 2023, \mn@doi [Phys. Rev. Lett.] {10.1103/PhysRevLett.131.171001}, 131, 171001

\bibitem[\protect\citeauthoryear{Smith, Poulin  \& Amin}{Smith et~al.}{2020}]{Axiclass_2}
Smith T.~L.,  Poulin V.,   Amin M.~A.,  2020, \mn@doi [Phys. Rev. D] {10.1103/PhysRevD.101.063523}, 101, 063523

\bibitem[\protect\citeauthoryear{Spergel et~al.,}{Spergel et~al.}{2015}]{Roman}
Spergel D.,  et~al., 2015, Wide-Field InfrarRed Survey Telescope-Astrophysics Focused Telescope Assets WFIRST-AFTA 2015 Report (\mn@eprint {arXiv} {1503.03757})

\bibitem[\protect\citeauthoryear{{Springel} et~al.,}{{Springel} et~al.}{2018}]{Springel:2018}
{Springel} V.,  et~al., 2018, \mn@doi [\mnras] {10.1093/mnras/stx3304}, \href {https://ui.adsabs.harvard.edu/abs/2018MNRAS.475..676S} {475, 676}

\bibitem[\protect\citeauthoryear{{Spurio Mancini}, {Piras}, {Alsing}, {Joachimi}  \& {Hobson}}{{Spurio Mancini} et~al.}{2022}]{cosmopower}
{Spurio Mancini} A.,  {Piras} D.,  {Alsing} J.,  {Joachimi} B.,   {Hobson} M.~P.,  2022, \mn@doi [\mnras] {10.1093/mnras/stac064}, \href {https://ui.adsabs.harvard.edu/abs/2022MNRAS.511.1771S} {511, 1771}

\bibitem[\protect\citeauthoryear{{Tegmark} \& {Zaldarriaga}}{{Tegmark} \& {Zaldarriaga}}{2002}]{Tegmark_Zaldarriaga}
{Tegmark} M.,  {Zaldarriaga} M.,  2002, \mn@doi [\prd] {10.1103/PhysRevD.66.103508}, \href {https://ui.adsabs.harvard.edu/abs/2002PhRvD..66j3508T} {66, 103508}

\bibitem[\protect\citeauthoryear{{The CMB-HD Collaboration} et~al.,}{{The CMB-HD Collaboration} et~al.}{2022}]{CMB_HD_snowmass}
{The CMB-HD Collaboration} et~al., 2022, \mn@doi [arXiv e-prints] {10.48550/arXiv.2203.05728}, \href {https://ui.adsabs.harvard.edu/abs/2022arXiv220305728T} {p. arXiv:2203.05728}

\bibitem[\protect\citeauthoryear{{The LSST Dark Energy Science Collaboration} et~al.,}{{The LSST Dark Energy Science Collaboration} et~al.}{2018}]{DESC_requirements}
{The LSST Dark Energy Science Collaboration} et~al., 2018, \mn@doi [arXiv e-prints] {10.48550/arXiv.1809.01669}, \href {https://ui.adsabs.harvard.edu/abs/2018arXiv180901669T} {p. arXiv:1809.01669}

\bibitem[\protect\citeauthoryear{{The Simons Observatory Collaboration} et~al.,}{{The Simons Observatory Collaboration} et~al.}{2025}]{SimonsObservatory:2025wwn}
{The Simons Observatory Collaboration} et~al., 2025, \mn@doi [arXiv e-prints] {10.48550/arXiv.2503.00636}, \href {https://ui.adsabs.harvard.edu/abs/2025arXiv250300636T} {p. arXiv:2503.00636}

\bibitem[\protect\citeauthoryear{{Villaescusa-Navarro}, {Marulli}, {Viel}, {Branchini}, {Castorina}, {Sefusatti}  \& {Saito}}{{Villaescusa-Navarro} et~al.}{2014}]{Villaescusa-Navarro_1}
{Villaescusa-Navarro} F.,  {Marulli} F.,  {Viel} M.,  {Branchini} E.,  {Castorina} E.,  {Sefusatti} E.,   {Saito} S.,  2014, \mn@doi [\jcap] {10.1088/1475-7516/2014/03/011}, \href {https://ui.adsabs.harvard.edu/abs/2014JCAP...03..011V} {2014, 011}

\bibitem[\protect\citeauthoryear{{Visinelli} \& {Vagnozzi}}{{Visinelli} \& {Vagnozzi}}{2019}]{Visinelli_2019}
{Visinelli} L.,  {Vagnozzi} S.,  2019, \mn@doi [\prd] {10.1103/PhysRevD.99.063517}, \href {https://ui.adsabs.harvard.edu/abs/2019PhRvD..99f3517V} {99, 063517}

\bibitem[\protect\citeauthoryear{{Vogt}, {Marsh}  \& {Lagu{\"e}}}{{Vogt} et~al.}{2023}]{vogt}
{Vogt} S. M.~L.,  {Marsh} D. J.~E.,   {Lagu{\"e}} A.,  2023, \mn@doi [\prd] {10.1103/PhysRevD.107.063526}, \href {https://ui.adsabs.harvard.edu/abs/2023PhRvD.107f3526V} {107, 063526}

\bibitem[\protect\citeauthoryear{{Wang} et~al.,}{{Wang} et~al.}{2022}]{Roman_forecast}
{Wang} Y.,  et~al., 2022, \mn@doi [\apj] {10.3847/1538-4357/ac4973}, \href {https://ui.adsabs.harvard.edu/abs/2022ApJ...928....1W} {928, 1}

\bibitem[\protect\citeauthoryear{Weinberg}{Weinberg}{1978}]{Weinberg_axion}
Weinberg S.,  1978, \mn@doi [Phys. Rev. Lett.] {10.1103/PhysRevLett.40.223}, 40, 223

\bibitem[\protect\citeauthoryear{{Widrow} \& {Kaiser}}{{Widrow} \& {Kaiser}}{1993}]{Widrow_schrodinger}
{Widrow} L.~M.,  {Kaiser} N.,  1993, \mn@doi [\apjl] {10.1086/187073}, \href {https://ui.adsabs.harvard.edu/abs/1993ApJ...416L..71W} {416, L71}

\bibitem[\protect\citeauthoryear{Wilczek}{Wilczek}{1978}]{Wilczek_1978}
Wilczek F.,  1978, \mn@doi [Phys. Rev. Lett.] {10.1103/PhysRevLett.40.279}, 40, 279

\bibitem[\protect\citeauthoryear{{Winch}, {Rogers}, {Hlo{\v{z}}ek}  \& {Marsh}}{{Winch} et~al.}{2024}]{Winch_Rogers_etal}
{Winch} H.,  {Rogers} K.~K.,  {Hlo{\v{z}}ek} R.,   {Marsh} D. J.~E.,  2024, \mn@doi [\apj] {10.3847/1538-4357/ad7a73}, \href {https://ui.adsabs.harvard.edu/abs/2024ApJ...976...40W} {976, 40}

\bibitem[\protect\citeauthoryear{Witten}{Witten}{1984}]{Witten_string_axion}
Witten E.,  1984, \mn@doi [Phys. Lett. B] {10.1016/0370-2693(84)90422-2}, 149, 351

\bibitem[\protect\citeauthoryear{{Wright} et~al.,}{{Wright} et~al.}{2025}]{KIDS_LEGACY}
{Wright} A.~H.,  et~al., 2025, \mn@doi [arXiv e-prints] {10.48550/arXiv.2503.19441}, \href {https://ui.adsabs.harvard.edu/abs/2025arXiv250319441W} {p. arXiv:2503.19441}

\bibitem[\protect\citeauthoryear{{Ye}, {Jiang}  \& {Silvestri}}{{Ye} et~al.}{2024}]{Silvestri}
{Ye} G.,  {Jiang} J.-Q.,   {Silvestri} A.,  2024, \mn@doi [arXiv e-prints] {10.48550/arXiv.2411.07082}, \href {https://ui.adsabs.harvard.edu/abs/2024arXiv241107082Y} {p. arXiv:2411.07082}

\bibitem[\protect\citeauthoryear{{Zimmermann}, {Marsh}, {Rogers}, {Winther}  \& {Shen}}{{Zimmermann} et~al.}{2024}]{Zimmerman_FDM}
{Zimmermann} T.,  {Marsh} D. J.~E.,  {Rogers} K.~K.,  {Winther} H.~A.,   {Shen} S.,  2024, \mn@doi [arXiv e-prints] {10.48550/arXiv.2412.10829}, \href {https://ui.adsabs.harvard.edu/abs/2024arXiv241210829Z} {p. arXiv:2412.10829}

\bibitem[\protect\citeauthoryear{{Zuntz} et~al.,}{{Zuntz} et~al.}{2015}]{Zuntz:2015}
{Zuntz} J.,  et~al., 2015, \mn@doi [Astronomy and Computing] {10.1016/j.ascom.2015.05.005}, \href {https://ui.adsabs.harvard.edu/abs/2015A&C....12...45Z} {12, 45}

\bibitem[\protect\citeauthoryear{{van Daalen}, {Schaye}, {Booth}  \& {Dalla Vecchia}}{{van Daalen} et~al.}{2011}]{vanDaalen:2011}
{van Daalen} M.~P.,  {Schaye} J.,  {Booth} C.~M.,   {Dalla Vecchia} C.,  2011, \mn@doi [\mnras] {10.1111/j.1365-2966.2011.18981.x}, \href {https://ui.adsabs.harvard.edu/abs/2011MNRAS.415.3649V} {415, 3649}

\bibitem[\protect\citeauthoryear{{van Daalen}, {McCarthy}  \& {Schaye}}{{van Daalen} et~al.}{2020}]{vandaalen:2020}
{van Daalen} M.~P.,  {McCarthy} I.~G.,   {Schaye} J.,  2020, \mn@doi [\mnras] {10.1093/mnras/stz3199}, \href {https://ui.adsabs.harvard.edu/abs/2020MNRAS.491.2424V} {491, 2424}

\makeatother
\end{thebibliography}

\appendix
\section{The effects of axions and baryonic feedback on the shear correlation function}\label{sec:app_A}
\begin{figure*}
    \begin{minipage}{\textwidth}
\includegraphics[width=1.\columnwidth]{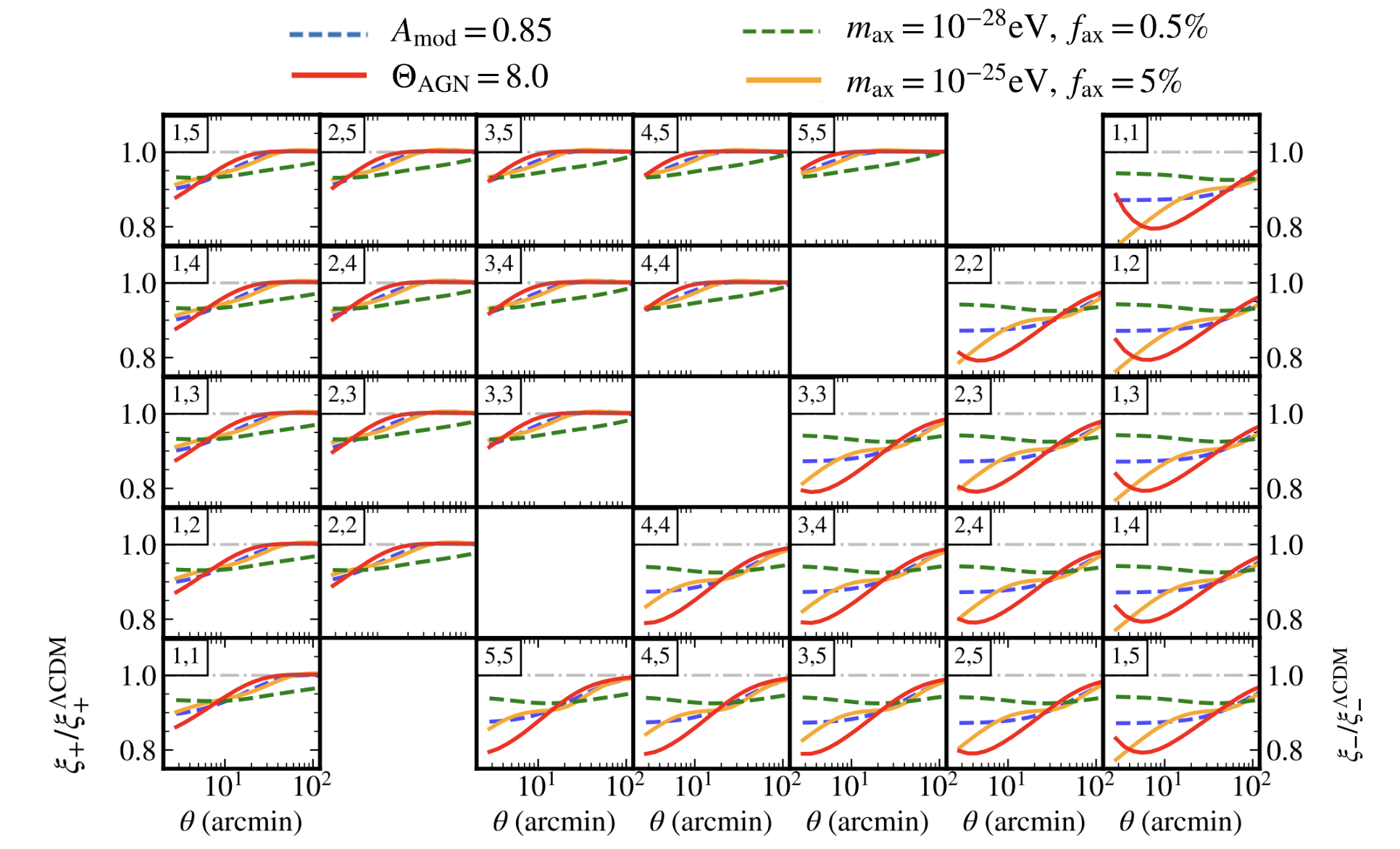} 
    \end{minipage}
\caption{The shear correlation function \(\xi_{+/-}(\theta)\) (\textit{left}/\textit{right}) of the power suppression scenarios introduced in Fig.~\ref{fig:toy_example} in ratio to the feedback-free \LCDM limit \(\xi_{+/-}^{\Lambda\mathrm{CDM}}(\theta)\), as a function of angular separation \(\theta\) for the five forecast LSST Y1 tomographic redshift bins (1 to 5, increasing in redshift; see Sec.~\ref{sec:data}) and their cross-correlations. We show \(\xi_{+/-}(\theta)\) for the \(\theta\) that can be accessed by LSST Y1; the forecast LSST Y1 data uncertainties are much smaller than the differences between models.}
\label{fig:correlation_functions}
\end{figure*}

In Fig.~\ref{fig:correlation_functions}, we show the shear correlation function of the power suppression scenarios introduced in Fig.~\ref{fig:toy_example} in ratio to the feedback-free \(\Lambda\)CDM limit. The effect of the power suppression is stronger in $\xi_{-}(\theta)$, which is more sensitive to smaller scales, although it is present also in $\xi_{+}(\theta)$. As with the matter power spectrum reconstruction in Sec.~\ref{sec:reconstruction_forecast}, the four scenarios are distinct also in the shear correlation function. Smaller angular separations are typically more suppressed. Because of the coupling between different wavenumbers in the shear correlation, this suppression is less clear than when reconstructing the matter power.

\section{Extended cosmological analysis: full posterior distributions of cosmological, axion and feedback parameters}\label{sec:app_B}
\begin{figure*}
\includegraphics[width=1.\textwidth]{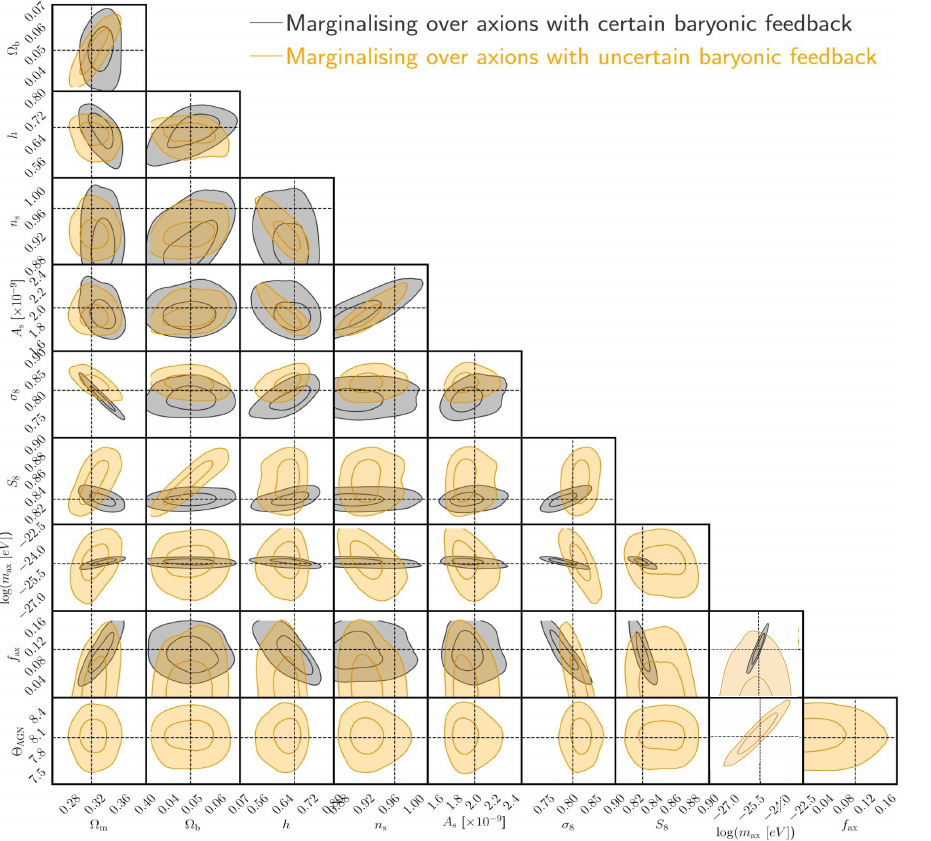} 
\caption{As the centre panel of Fig.~\ref{fig:axion_feedback_degeneracy} except for all \(\Lambda\)CDM, axion and feedback parameters (see Table \ref{tab:priors}), as well as the derived parameters \(\sigma_8\) and \(S_8\).}
\label{fig:full_posterior}
\end{figure*}

In Fig.~\ref{fig:full_posterior}, we show the full posteriors of cosmological, axion and feedback parameters for the two inferences shown in the centre panel of Fig.~\ref{fig:axion_feedback_degeneracy}. We do not show the systematics parameters (Table \ref{tab:priors}) although these are marginalised; there are no strong degeneracies between axion and systematics parameters. There are degeneracies between axion parameters and $S_8$ as anticipated by \citet{Rogers_S8tensionaxions}. Higher \(f_\mathrm{ax}\) decreases \(\sigma_8\), which in turn lowers \(S_8\). It is beyond the scope of this work to assess the full prospects for addressing the so-called ``\(S_8\) tension.''

\end{document}